\documentclass[showpacs,preprintnumbers,amsmath,amssymb,twocolumn,superscriptaddress,aps,floatfix]{revtex4}

\usepackage{graphicx}
\usepackage{dcolumn}
\usepackage{bm}
\usepackage{textcomp}
\usepackage{amsmath}
\usepackage{subfigure}

\newcommand{\sss}{\scriptscriptstyle}
\newcommand{\sst}{\scriptstyle}

\newcommand{\stext}[1]{\sss \text{#1} \sst}

\renewcommand{\emph}[1]{\textit{#1}}

\begin{document}
\title{Anisotropy, phonon modes, and free charge carrier parameters in monoclinic $\beta$-gallium oxide single crystals}

\author{M.~Schubert}
\email{schubert@engr.unl.edu}
\homepage{http://ellipsometry.unl.edu}
\affiliation{Department of Electrical Engineering and Center for Nanohybrid Functional Materials, University of Nebraska-Lincoln, U.S.A.}
\affiliation{Leibniz Institute for Polymer Research, Dresden, Germany}
\author{R.~Korlacki}
\affiliation{Department of Electrical Engineering and Center for Nanohybrid Functional Materials, University of Nebraska-Lincoln, U.S.A.}
\author{S.~Knight}
\affiliation{Department of Electrical Engineering and Center for Nanohybrid Functional Materials, University of Nebraska-Lincoln, U.S.A.}
\author{T.~Hofmann}
\affiliation{Department of Electrical Engineering and Center for Nanohybrid Functional Materials, University of Nebraska-Lincoln, U.S.A.}
\affiliation{\mbox{Department of Physics, Chemistry and Biology, IFM, Link\"{o}ping University, SE-581 83 Link\"{o}ping, Sweden}}
\author{S.~Sch\"{o}che}
\affiliation{J.~A.~Woollam Corporation, Inc.}
\author{V.~Darakchieva}
\affiliation{\mbox{Department of Physics, Chemistry and Biology, IFM, Link\"{o}ping University, SE-581 83 Link\"{o}ping, Sweden}}
\author{E.~Janz$\acute{e}$n}
\affiliation{\mbox{Department of Physics, Chemistry and Biology, IFM, Link\"{o}ping University, SE-581 83 Link\"{o}ping, Sweden}}
\author{B.~Monemar}
\affiliation{\mbox{Department of Physics, Chemistry and Biology, IFM, Link\"{o}ping University, SE-581 83 Link\"{o}ping, Sweden}}
\affiliation{\mbox{Global Innovation Research Organization, Tokyo University of Agriculture and Technology, Koganei, Tokyo, Japan}}
\author{D. Gogova}
\affiliation{\mbox{Central Laboratory of Solar Energy and New Energy Sources, Bulgarian Academy of Sciences, Sofia, Bulgaria}}
\affiliation{\mbox{Leibniz Institute for Crystal Growth, Berlin, Germany}}
\author{Q.-T.~Thieu}
\affiliation{\mbox{Global Innovation Research Organization, Tokyo University of Agriculture and Technology, Koganei, Tokyo, Japan}}
\affiliation{\mbox{Department of Applied Chemistry, Tokyo University of Agriculture and Technology, Koganei, Tokyo, Japan}}
\author{R.~Togashi}
\affiliation{\mbox{Department of Applied Chemistry, Tokyo University of Agriculture and Technology, Koganei, Tokyo, Japan}}
\author{H.~Murakami}
\affiliation{\mbox{Department of Applied Chemistry, Tokyo University of Agriculture and Technology, Koganei, Tokyo, Japan}}
\author{Y.~Kumagai}
\affiliation{\mbox{Department of Applied Chemistry, Tokyo University of Agriculture and Technology, Koganei, Tokyo, Japan}}
\author{K.~Goto}
\affiliation{\mbox{Department of Applied Chemistry, Tokyo University of Agriculture and Technology, Koganei, Tokyo, Japan}}
\affiliation{\mbox{Tamura Corporation, Sayama, Saitama, Japan}}
\author{A.~Kuramata}
\affiliation{\mbox{Tamura Corporation, Sayama, Saitama, Japan}}
\author{S.~Yamakoshi}
\affiliation{\mbox{Tamura Corporation, Sayama, Saitama, Japan}}
\author{M.~Higashiwaki}
\affiliation{\mbox{National Institute of Information and Communications Technology, Koganei, Tokyo, Japan}}

\date{}

\begin{abstract}
We derive a dielectric function tensor model approach to render the optical response of monoclinic and triclinic symmetry materials with multiple uncoupled infrared and farinfrared active modes. We apply our model approach to monoclinic $\beta$-Ga$_2$O$_3$ single crystal samples. Surfaces cut under different angles from a bulk crystal, (010) and ($\bar{2}$01), are investigated by generalized spectroscopic ellipsometry within infrared and farinfrared spectral regions. We determine the frequency dependence of 4 independent $\beta$-Ga$_2$O$_3$ Cartesian dielectric function tensor elements by matching large sets of experimental data using a point by point data inversion approach. From matching our monoclinic model to the obtained 4 dielectric function tensor components, we determine all infared and farinfrared active transverse optic phonon modes with $A_u$ and $B_u$ symmetry, and their eigenvectors within the monoclinic lattice. We find excellent agreement between our model results and results of density functional theory calculations. We derive and discuss the frequencies of longitudinal optical phonons in $\beta$-Ga$_2$O$_3$. We derive and report density and anisotropic mobility parameters of the free charge carriers within the tin doped crystals. We discuss the occurrence of longitudinal phonon plasmon coupled modes in $\beta$-Ga$_2$O$_3$ and provide their frequencies and eigenvectors. We also discuss and present monoclinic dielectric constants for static electric fields and frequencies above the reststrahlen range, and we provide a generalization of the Lyddane-Sachs-Teller relation for monoclinic lattices with infrared and farinfrared active modes. We find that the generalized Lyddane-Sachs-Teller relation is fulfilled excellently for $\beta$-Ga$_2$O$_3$. 
\end{abstract}
 \maketitle

\section {Introduction}

Group-III sesquioxides have regained interest as wide band gap semiconductors with unexploited physical properties. Electric conductivity in transparent, polycrystalline, tin doped In$_2$O$_3$ and Ga$_2$O$_3$ facilitates thin film electrodes for “smart windows”~\cite{GranqvistBook1995,GogovaJCG1999}, photovoltaics~\cite{GranqvistBook1995}, large area flat panel displays~\cite{BetzSCT2006}, and sensors, for example~\cite{ RetiSA1994}. The highly anisotropic monoclinic $\beta$-gallia crystal structure ($\beta$ phase) is the most stable crystal structure among the five phases ($\alpha$, $\beta$, $\gamma$, $\varepsilon$, and $\delta$) of Ga$_2$O$_3$~\cite{RoyJACS1952,TippinsPR1965}. Mixed phase $\alpha-\beta$ Ga$_2$O$_3$ oxide junctions were recently discovered for high activity photocatalytic water splitting~\cite{Ming-GangJMCA2014}. Current research focusses on the development of single crystalline group-III sesquioxide semiconductors with low defect densities for potential use as active materials in electronic and optoelectronic devices~\cite{UeadaAPL1997Ga2O3}. The thermodynamically stable $\beta$-Ga$_2$O$_3$ phase is of particular interest due to its large band gap energy of 4.85~eV, lending promise for applications in short wavelength photonics and transparent electronics~\cite{WagnerScience2003}. The high electric break down field value of $\beta$-Ga$_2$O$_3$, which is estimated at 8 MVcm$^{-1}$ exceeds those of contemporary semiconductor materials such as Si, GaAs, SiC, group-III nitrides, or ZnO~\cite{KoheiJCG2013}. Baliga's figure of merit for $\beta$-Ga$_2$O$_3$ is several times larger than those for 4H-SiC or GaN~\cite{BaligaIEEEEDL1989}. Baliga's figure of merit is the basic parameter to evaluate a material's suitability for power device applications. The figure of merit is proportional to the cube of the breakdown field, but only linearly proportional to mobility, hence, a large breakdown field can trump small mobility. Melt growth methods of bulk single crystals have been demonstrated by Czochralski growth~\cite{TommJCG2000}, float zone growth~\cite{VilloraJCG2004}, and edge-defined film-fed growth~\cite{Higashiwakipssa2014} suitable for mass production due to cost efficiency compared with growth of GaN substrates, for example~\cite{GogovaJAP2013}. Homoepitaxial thin film growth was developed by molecular beam epitaxy~\cite{KoheiJCG2013}, and metal-organic vapor phase epitaxy methods~\cite{GogovaCEC2015} yielding good quality crystalline materials. Schottky barrier diodes (SBD) and metal-semiconductor field-effect transistors (MESFETs) on $\beta$-Ga$_2$O$_3$ homoepitaxial layers were reported for the first time by Sasaki~\textit{et al.}~\cite{KoheiJCG2013} and a breakdown voltage of 125 V was obtained. The MESFETs also exhibited excellent characteristics such as a nearly ideal pinch-off of the drain current, an off-state breakdown voltage over 250 V, a high on/off drain current ratio of around 10$^{4}$, and small gate leakage current~\cite{Higashiwakipssa2014}. These device characteristics clearly indicate the great potential of $\beta$-Ga$_2$O$_3$ as a high power device material. It is also expected that extremely wide band gap semiconductors (with band gap energies larger than 4~eV) may have potential for so far unexplored optoelectronic applications in the deep ultra violet region. Such applications are emerging in the biotechnology and nanotechnology areas. For example, combining scanning near field optical microscopy~\cite{BetzigScience1991} with deep ultra violet transparent optical fibers~\cite{OtoIEEEPT2001} may enable imaging of molecular structures of DNA and proteins using characteristic absorption and/or fluorescence. Rare earth or 3d transition metal doping in $\beta$-Ga$_2$O$_3$ thin films further demonstrated promising optical and photoluminescent properties, for example in thin film electroluminescent devices~\cite{ShinoyaPhosphorHandbook,MiyataJL2000}.

Crucial for device design and operation is knowledge on electrical transport parameters. Likewise, understanding of heat transport as well as phonon assisted free charge carrier scattering requires precise knowledge on long wavelength phonon energies and band structure properties. In this paper we investigate lattice and free charge carrier properties of $\beta$-Ga$_2$O$_3$ by experiment and by calculation of phonon mode parameters. Knowledge on phonon modes and free charge carrier parameters is not exhaustive for $\beta$-Ga$_2$O$_3$. Very few reports exist on experimental determination of phonon mode parameters and their anisotropy~\cite{Villorapssa2002Ga2O3IR}. No report exists to our best knowledge which observes and describes coupling of phonon and free charge carrier modes. Few reports exist on theoretical prediction and experimental determination of static and high frequency dielectric constants and their anisotropy~\cite{PasslackJAP1995Ga2O3film,PasslackAPL1994Ga2O3films,HoeneisenSSE1971Ga2O3DC,HePRB2006Ga2O3calc,SchmitzJAP1998Ga2O3films,RebienAPL2002Ga2O3SEfilms,LiuAPL2007Ga2O3phononcalc}. Calculations predict effective mass parameters~\cite{UeadaAPL1997Ga2O3,Peelaerspssb2015Ga2O3meff,JuJMCA2014Ga2O3theory,YamaguchiSSC2004FPGa2O3,HeAPL2006Ga2O3theory}, and few experiments were reported~\cite{MohamedAPL2010Ga2O3HRARPES,JanowitzNJP2011}. Theoretical descriptions of Brilloun zone center phonon modes are reported~\cite{LiuAPL2007Ga2O3phononcalc}, and phonon band structures and density of states allowed prediction of thermal transport properties. Recently, Guo \textit{et al}. measured the thermal conductivity in $\beta$-Ga$_2$O$_3$ single crystals and observed behaviors indicative for phonon assisted heat transport with strongly anisotropic group velocities supported by first principles calculations~\cite{GuoAPL2015thcondGa2O3}.

Owing to the unique strength of ellipsometry to resolve the state of polarization of light reflected off or transmitted through samples, both real and imaginary parts of the complex dielectric function can be determined at optical wavelengths~\cite{Drude87,Drude88,Drude_1900}. Generalized ellipsometry extends this concept to arbitrarily anisotropic materials, and allows to determine, in principle, all 9 complex-valued elements of the dielectric function tensor~\cite{SchubertADP15_2006}. Jellison \textit{et al.} reported generalized ellipsometry analysis of a monoclinic crystal, CdWO$_4$~\cite{JellisonPRB2011CdWO4}. Experimental data were taken from multiple sample orientations in the near infrared to ultra violet spectral regions. It was shown that 4 complex-valued dielectric tensor elements are required for each wavelength, which were determined spectroscopically, and independently of physical model line shape functions. The authors pointed out that no general rotations could be found to diagonalize the 4 tensor elements independently of wavelength. In the transparency region, a diagonalization could be found, but only one which depends on wavelengths. Jellison~\textit{et al.} suggested to record and present, in general for monoclinic materials, 4 instead of 3 independent spectroscopic dielectric function tensor elements. In this context, a 4$^{th}$ spectroscopic response function is described, whose physical meaning, however, remained unexplained. Kuz'menko \textit{et al.} and M\"{o}ller \textit{et al.} analyzed polarized reflectance from multiple surface orientations of monoclinic crystals, CuO and MnWO$_4$, respectively~\cite{KuzmenkoPRB2001CuOIR,MoellerPRB2014MnWO4}. Spectra were obtained as a function of incident light polarization relative to the crystallographic axes. The authors used a physical function lineshape model first described by Born and Huang~\cite{Born54}. This lineshape model brings 4 interdependent dielectric function tensor elements into existence for monoclinic materials. Kuz'menko~\cite{KuzmenkoJPCM1996aBi2O3} described this model in more detail and exemplified  analysis of the partially polarization resolved reflectance spectra for monoclinic $\alpha$-Bi$_2$O$_3$. The Born and Huang model allows for the derivation of TO modes and their unit eigen displacement vectors. These were obtained and reported in Refs.~\cite{KuzmenkoPRB2001CuOIR,MoellerPRB2014MnWO4,KuzmenkoJPCM1996aBi2O3}. However, numerical integrations were required to guarantee Kramers-Kronig consistency for the 4 tensor element spectra, since neither of these elements can be obtained independently and as complex-valued functions from reflectance data analysis. To our best knowledge, no independent verification of the Born and Huang model was provided for monoclinic crystals, where the dielectric tensor element functions have been determined independently and without physical lineshape functions. Furthermore, the determination of longitudinal optical modes as well as plasma coupling in crystals with monoclinic symmetry has not been discussed and presented within the Born and Huang model. Also, the Lyddane-Sachs-Teller relation is not valid for monoclinic lattices and we present its generalization in this paper. We apply our model to $\beta$-Ga$_2$O$_3$ single crystals, and obtain and discuss fundamental physical parameters for this potentially important semiconductor material.

\begin{figure}[!tbp]
  \begin{center}
    \includegraphics[width=0.3\linewidth]{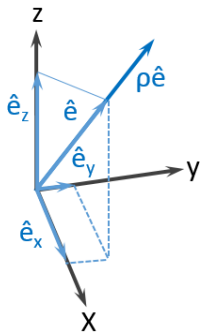}
    \caption{Unit eigen displacement vector $\hat{e}$ characteristic for a dielectric eigen polarizability $\mathbf{P}_{\hat{e}}$ whose frequency response is rendered by a complex-valued response function $\varrho_{\hat{e}}$.}
    \label{fig:eigendisplacement}
  \end{center}
\end{figure}

\section{Theory}

The lattice constants of $\beta$-Ga$_2$O$_3$ are $a=12.23$ \AA, $b=3.04$ \AA, and $c=5.80$ \AA, and the monoclinic angle is $\beta= 103.7^{\circ}$~\cite{GellerJPC1960Ga2O3}. There are ten atoms in the primitive unit cell of $\beta$-Ga$_2$O$_3$ with 30 normal modes of vibrations. The irreducible representation for acoustical and optical zone center modes are: $\Gamma_{\mathrm{aco}}=A_u+2B_u$ and $\Gamma_{\mathrm{opt}}= 10A_g+ 4A_u+ 5B_g+8B_u$. For the optical modes, $A_g$ and $B_g$ modes are Raman active, while $A_u$ and $B_u$ modes are long wavelength (infrared and farinfrared) active. Hence, $\beta$-Ga$_2$O$_3$ is a material with multiple modes of long wavelength active phonons and plasmons. We provide a simple approach to construct the dielectric function tensor of materials with non orthogonal normal modes. Born and Huang provided both an atomistic as well as a microscopic description of the lattice dynamics at long wavelengths from first principles and elasticity theory~\cite{Born54}. Both approaches lead to a description of the dielectric function tensor to which the result of our approach is equivalent. While our approach is straightforward, we extend the Born and Huang model by discussion of non orthogonal longitudinal optical modes and coupling with plasma modes. All normal modes with transverse and longitudinal character predicted by theory are observed in our experiment and will be discussed in detail.

\subsection{Uncoupled Eigen Polarizability Model}
\label{sec:AnisDFTensor}

Intrinsic dielectric polarizations (eigen displacement modes) of a homogeneous material give rise to long wavelength active phonon modes. Each mode is associated with an electric dipole charge oscillation. The dipole axis can be associated with a characteristic eigenvector (unit eigen displacement vector $\hat{e}$). Within the frequency domain, and within a Cartesian system with unit directions \textbf{x}, \textbf{y}, \textbf{z}, the dielectric polarizability $\mathbf{P}$ under the influence of an electric phasor field $\mathbf{E}$ along $\mathbf{\hat{e}}=\hat{e}_x\mathbf{x}+\hat{e}_y\mathbf{y}+\hat{e}_z\mathbf{z}$ is then given by a complex-valued response function $\varrho_{\hat{e}}$ (Fig.~\ref{fig:eigendisplacement})

\begin{equation}
\mathbf{P}_{\hat{e}}=\varrho_{\hat{e}}(\mathbf{\hat{e}}\mathbf{E})\mathbf{\hat{e}}.
\end{equation}

\noindent Function $\varrho_{\hat{e}}$ must satisfy causality and energy conservation requirements, i.e., the Kramers-Kronig integral relations and $Im \{ \varrho_{\hat{e}} \} \ge 0, \forall$ $\omega \ge 0$~\cite{Dressel_2002,Jackson75}. Under the assumption that different eigen displacement modes do not couple, their eigenvectors may lie along certain, fixed spatial directions within a given sample of material. The linear polarization response of a material with $m$ eigen displacement modes is then obtained from summation

\begin{equation}\label{eq:sumP}
\mathbf{P}=\sum^{m}_{l=1}\mathbf{P}_{\hat{e_l}}=\sum^{m}_{l=1}\varrho_{\hat{e}_l}(\mathbf{\hat{e}}_l\otimes\mathbf{\hat{e}}_l)\mathbf{E}= \chi\mathbf{E},
\end{equation}

\noindent where $\otimes$ is the dyadic product. Eq.~(\ref{eq:sumP}) results in a dielectric polarization response tensor $\chi$, which is fully symmetric in all indices

\begin{equation}\label{eq:xhiij}
(\chi)_{ij}=\sum^{m}_{l=1}\varrho_{\hat{e}_l}\hat{e}_{i,l}\hat{e}_{j,l}=(\chi)_{ji},\mbox{}i,j=``x",``y",``z".
\end{equation}

\noindent The mutual orientations of the eigenvectors, and the frequency responses of their eigen displacements determine the optical character of a given, dielectrically polarizable material. For certain or all frequency regions, analogies can be found with symmetry properties of monoclinic, triclinic, orthorhombic, tetragonal, hexagonal, trigonal, or cubic crystal classes. The field phasors displacement $\mathbf{D}$, and $\mathbf{E}$ are related by the dielectric function tensor ($\varepsilon_0$ is the vacuum permittivity)

\begin{equation}\label{eq:eps}
\mathbf{D}=\varepsilon_0\left(1+\chi\mathbf{E}\right)=\varepsilon_0\varepsilon\mathbf{E}.
\end{equation}

Likewise to $\chi$, $\varepsilon$ is fully symmetric, invariant under time and space inversion, and a function of frequency $\omega$. Chiral arrangements of eigen displacements require augmentation of coupling between eigen modes, which is not further discussed here. The dielectric function tensor in Eq.~(\ref{eq:eps}) has 6 independent complex-valued parameters. These render physical observables, which can be obtained by experiment, for example using generalized spectroscopic ellipsometry~\cite{SchubertIRSEBook_2004}. The dielectric function tensor contains information on fundamental physical properties. For example, the frequencies of two characteristic optical modes, transverse optical (TO; $\omega_{\stext{TO}}$) and longitudinal optical (LO; $\omega_{\stext{LO}}$), can be obtained, respectively, from the roots of the determinants of $\varepsilon^{-1}$, and $\varepsilon$

\begin{equation}\label{eq:eps:TO}
0=\det\{ \varepsilon^{-1}(\omega_{\stext{TO}})\},
\end{equation}

\begin{equation}\label{eq:eps:LO}
0=\det\{\varepsilon(\omega_{\stext{LO}})\}.
\end{equation}

Each of the modes $\omega_{\stext{TO}}$ and $\omega_{\stext{LO}}$ are associated with a unit eigen displacement vector, $\mathbf{\hat{e}}_{\stext{TO}}$ and $\mathbf{\hat{e}}_{\stext{LO}}$, which can be obtained, respectively, from the set of equations

\begin{equation}\label{eq:eps:TOeigenvector}
0=\varepsilon^{-1}(\omega_{\stext{TO}})\mathbf{\hat{e}}_{\stext{TO}},
\end{equation}

\begin{equation}\label{eq:eps:LOeigenvector}
0=\varepsilon(\omega_{\stext{LO}})\mathbf{\hat{e}}_{\stext{LO}}.
\end{equation}

\subsection{Dielectric Function Tensor Model for $\beta$-Ga$_2$O$_3$}
\label{sec:DFTensorModel}

\begin{figure}[!tbp]
  \begin{center}
    \includegraphics[width=0.4\linewidth]{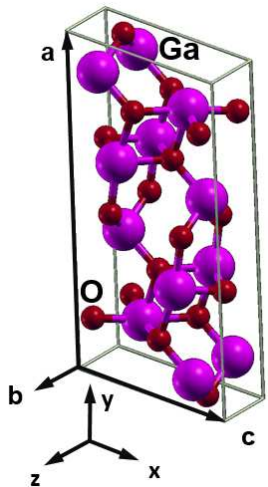}
    \caption{Unit cell of $\beta$-Ga$_2$O$_3$.}
    \label{fig:Ga2O3unitcell}
  \end{center}
\end{figure}

Long wavelength active phonons correspond to lattice displacements, which are associated with a linear dipole moment. In $\beta$-Ga$_2$O$_3$ (Fig.~\ref{fig:Ga2O3unitcell}), 12 long wavelength active phonon branches are predicted by symmetry.  Each branch consists of a pair of TO and LO modes. In the presence of free charge carriers, 3 additional LO modes occur due to 3 available dimensions for plasmon propagation. Their eigen displacement vectors have to be determined from experiment, as will be discussed further below. The free charge carrier modes couple with the LO modes of the phonon branches unless their eigen displacement vectors are orthogonal. This coupling leads to new experimentally observable modes, the so called longitudinal phonon plasmon (LPP; $\omega_{\stext{LPP}}$) modes. 

\paragraph{Transverse optical modes:} Modes with $A_u$ symmetry (4) are polarized along $\mathbf{b}$ only. Modes with $B_u$ symmetry are polarized within the $\mathbf{a}-\mathbf{c}$ plane (Fig.~\ref{fig:Ga2O3unitcell}). A choice of coordinates must be made at this step. We align unit cell axes $\mathbf{b}$ and $\mathbf{c}$ with $\mathbf{z}$ and $\mathbf{x}$, respectively, and $\mathbf{a}$ is within the (x-y) plane. We introduce vector $\mathbf{\hat{a}}$ parallel to $\mathbf{y}$ for convenience, and obtain $\mathbf{\hat{a}}$, $\mathbf{b}$, $\mathbf{c}$ as a pseudo orthorhombic system. Then, Eq.~(\ref{eq:sumP}) leads to the following summations

\begin{equation}\label{eq:bGa2O3polarization}
\mathbf{P^{\beta-Ga_2O_3}}= \sum^{8}_{j=1}\varrho^{B_u}_{j}(\cos\alpha_j\mathbf{x} + \sin\alpha_j\mathbf{y}) + \sum^{4}_{k=1}\varrho^{A_u}_{k}\mathbf{z},
\end{equation}

\noindent where $\alpha_j$ describes the dipole oscillation axis of the $j^{th}$ $B_u$ mode relative to $\mathbf{c}$. As a result, and within the chosen coordinate frame, the dielectric function tensor has 4 independent complex-valued elements: $\varepsilon_{xx}$, $\varepsilon_{xy}$, $\varepsilon_{yy}$, and $\varepsilon_{zz}$. 

The energy dependent contribution to the long wavelength polarization response of an uncoupled electric dipole charge oscillation is commonly described using a Lorentzian broadened oscillator function~\cite{SchubertIRSEBook_2004,HumlicekHOE}

\begin{equation}\label{eq:rho}
\varrho_{(l)} \left(\omega\right)=\frac{A_{(l)}}{\omega^2_{\stext{TO},(l)}-\omega^2-i\omega\gamma_{(l)}},
\end{equation}

\noindent where $A_{(l)}$, $\omega_{\stext{TO},(l)}$, and $\gamma_{(l)}$ denote the amplitude, resonance frequency, and broadening parameter of a lattice resonance with TO character, $\omega$ is the frequency of the driving electromagnetic field, and $i=\sqrt{-1}$ is the imaginary unit. The index $l$ numerates the contributions of all independent dipole oscillations. 

\paragraph{Free charge carrier contributions:}  The energy dependent contribution to the long wavelength polarization response of free charge carriers is commonly described using the Drude model function~\cite{Kittel86,SchubertIRSEBook_2004,Fujiwara_2007,Pidgeon80}

\begin{equation}\label{eq:Drude}
\varrho_{\mathrm{\stext{FCC}},(x,y,z)} = -
\frac{\mathrm{e}^2 N}{\tilde{\varepsilon }_0 m_{\stext{eff},(x,y,z)} \omega (\omega + \mathrm{i}\gamma _{\stext{p},(x,y,z)} )},
\end{equation}
where $N$ is the free charge carrier volume density parameter. As discussed further below, we find the eigen displacement vectors of the plasma modes orthogonal to each other, and we cast their contributions within the  choice of Cartesian coordinates ($\mathbf{x},\mathbf{y},\mathbf{z}$) shown in Fig.~\ref{fig:Ga2O3unitcell}. Hence, the effective mass and plasma broadening parameters, $m_{\stext{eff},(x,y,z)}$ and $\gamma _{\stext{p},(x,y,z)}$, are indicated by their Cartesian axes, respectively ($\tilde {\varepsilon }_0$ is the vacuum permittivity, and {\it e} is the amount of the electrical unit charge). The plasmon broadening parameters can be related to optical mobility parameters $\mu_{(x,y,z)}$
\begin{equation}\label{eq:gamma-plasma}
\gamma_{\stext{p},(x,y,z)} = \frac{e}{m_{\stext{eff},(x,y,z)} \mu_{(x,y,z)} }.
\end{equation}

\paragraph{High frequency dielectric constants:} Equations (\ref{eq:rho}) and (\ref{eq:Drude}) vanish for large frequencies, however, contributions to the polarization functions may arise from higher frequency charge oscillations such as electronic band to band transitions. A full analysis requires the incorporation of experimental data far into the ultra violet region to identify the eigen displacement vectors of the electronic band to band transitions in $\beta$-Ga$_2$O$_3$. Because the fundamental band to band transition energy is far outside the spectral range investigated here, we approximate the high frequency contributions by frequency independent parameters which represent the sum of all contributions from all higher energy electronic band to band transitions

\begin{equation}\label{eq:monoclinicepsinf}
\boldsymbol{\varepsilon}_{\infty}= \left(
\begin{array}{ccc}
\varepsilon_{\infty,xx} & \varepsilon_{\infty,xy} & 0 \\
\varepsilon_{\infty,xy} & \varepsilon_{\infty,yy} & 0 \\
0           & 0 & \varepsilon_{\infty,zz}

\end{array}\right).
\end{equation}

\noindent Note that due to the monoclinic symmetry, 4 real-valued parameters are required. An effective eigen displacement vector can be found from Eq.~(\ref{eq:xhiij}) for the band gap spectral region, which may also be considered as effective monoclinic angle for this spectral region 

\begin{equation}\label{eq:monoclinicepsinfangle}
\alpha_{\infty} = \tan^{-1}\left(\frac{\varepsilon_{\infty,xy}}{\varepsilon_{\infty,xx}-1}\right)= \cot^{-1}\left(\frac{\varepsilon_{\infty,yy}-1}{\varepsilon_{\infty,xy}}\right).
\end{equation}

\paragraph{Static dielectric constants:} Equations (\ref{eq:rho}) contribute constant values at zero frequencies, when free charge carrier contributions in Eqs.~(\ref{eq:Drude}) are absent

\begin{equation}\label{eq:monoclinicepsDC}
\boldsymbol{\varepsilon}_{\stext{DC}}= \left(
\begin{array}{ccc}
\varepsilon_{\stext{DC},xx} & \varepsilon_{\stext{DC},xy} & 0 \\
\varepsilon_{\stext{DC},xy} & \varepsilon_{\stext{DC},yy} & 0 \\
0          & 0 & \varepsilon_{\stext{DC},zz}

\end{array}\right).
\end{equation}

\noindent The contributions are obtained explicitly as

\begin{equation}\label{eq:monoclinicepsxxDCangle}
\varepsilon_{\stext{DC},xx}= \varepsilon_{\infty,xx}+\sum^{8}_{j=1}\cos^2\alpha_j\frac{A^{B_u}_{j}}{\omega^2_{\stext{TO},j}},
\end{equation}

\begin{equation}\label{eq:monoclinicepsxyDC}
\varepsilon_{\stext{DC},xy}= \varepsilon_{\infty,xy}+\sum^{8}_{j=1}\sin\alpha_j\cos\alpha_j\frac{A^{B_u}_{j}}{\omega^2_{\stext{TO},j}},
\end{equation}

\begin{equation}\label{eq:monoclinicepsyyDC}
\varepsilon_{\stext{DC},yy}= \varepsilon_{\infty,yy}+\sum^{8}_{j=1}\sin^2\alpha_j\frac{A^{B_u}_{j}}{\omega^2_{\stext{TO},j}},
\end{equation}

\begin{equation}\label{eq:monoclinicepszzDC}
\varepsilon_{\stext{DC}, zz}= \varepsilon_{\infty,zz}+\sum^{4}_{k=1}\frac{A^{A_u}_{k}}{\omega^2_{\stext{TO},k}}.
\end{equation}

\noindent Hence, 4 constitutive parameters may be required near DC frequencies to describe the dielectric response of $\beta$-Ga$_2$O$_3$. An effective monoclinic eigen displacement vector within the $\mathbf{a}$-$\mathbf{c}$ plane can be found from Eq.~(\ref{eq:xhiij}), valid near DC frequencies only  

\begin{equation}\label{eq:monoclinicepsDCang}
\alpha_{\stext{DC}} = \tan^{-1}\left(\frac{\varepsilon_{\stext{DC},xy}}{\varepsilon_{\stext{DC},xx}-1}\right)= \cot^{-1}\left(\frac{\varepsilon_{\stext{DC},yy}-1}{\varepsilon_{\stext{DC},xy}}\right).
\end{equation}

\paragraph{Dielectric function tensor:} The $\beta$-Ga$_2$O$_3$ monoclinic dielectric function tensor is composed of the high frequency contributions, the dipole charge resonances, and the free charge carrier contributions

\begin{subequations}\label{eq:epsmonoall}
\begin{align}
\varepsilon_{xx} &= \varepsilon_{\infty,xx}+\sum^{8}_{j=1}\varrho^{B_u}_{j}\cos^2\alpha_j+\varrho_{\stext{FCC},x},\\
\varepsilon_{xy} &= \varepsilon_{\infty,xy}+\sum^{8}_{j=1}\varrho^{B_u}_{j}\sin\alpha_j\cos\alpha_j,\\
\varepsilon_{yy} &= \varepsilon_{\infty,yy}+\sum^{8}_{j=1}\varrho^{B_u}_{j}\sin^2\alpha_j+\varrho_{\stext{FCC},y},\\
\varepsilon_{zz} &= \varepsilon_{\infty,zz}+\sum^{4}_{k=1}\varrho^{A_u}_{k}+\varrho_{\stext{FCC},z},\\
\varepsilon_{xz} &= \varepsilon_{zx} =0.
\end{align}
\end{subequations}

Eqs.~\ref{eq:epsmonoall} provide valuable insight into the dielectric function tensor elements. If modes with $A_u$ and $B _u$ symmetry are distinct, critical point features~\cite{SchubertPRB61_2000} due to responses at frequencies with $A_u$ symmetry should only occur in $\varepsilon_{zz}$. Features due to modes with $B_u$ symmetry should only occur in $\varepsilon_{xx}$, $\varepsilon_{xy}$, and $\varepsilon_{yy}$. Depending on the orientation of the unit eigen displacement vector of a given mode, contributions may occur either (i) in $\varepsilon_{xx}$ ($\alpha=0^{\circ}$) only, or (ii) in $\varepsilon_{yy}$ ($\alpha=90^{\circ}$) only,  or (iii) in all $\varepsilon_{xx}$, $\varepsilon_{xy}$, and $\varepsilon_{yy}$ ($\alpha \ne n\pi$, $n=0,\pm 1, \pm 2, \dots$). Element $\varepsilon_{xy}$ is different from zero in case (iii) only. The imaginary part of $\varepsilon_{xy}$ can be negative. The latter provides a unique experimental access to identify whether $\alpha$ for a given mode shares an acute, a right, or an obtuse angle with the $\mathbf{c}$ axis. Note that $\varepsilon_{xx}$, $\varepsilon_{xy}$, and $\varepsilon_{yy}$ over determine the intrinsic polarizability functions. This is because $\varepsilon_{xy}$ is the product of simple geometrical shear projections and not the result of new, or additional physical properties in materials with non orthogonal unit eigen displacement vectors of intrinsic modes.

\paragraph{LO mode determination:} The determinant in Eq.~(\ref{eq:eps:LO}) factorizes into 2 equations, one valid for electric field polarization within the $x$-$y$ plane, and one equation valid for polarization along $z$, respectively,

\begin{equation} \label{eq:acLO}
0=\varepsilon_{xx}(\omega_{\stext{LO}_{(n)}})\varepsilon_{yy}(\omega_{\stext{LO}_{(n)}})-\varepsilon^2_{xy}(\omega_{\stext{LO}_{(n)}}).
\end{equation}

\noindent and

\begin{equation} \label{eq:bLO}
0=\varepsilon_{zz}(\omega_{\stext{LO}_{(n)}}).
\end{equation}

\noindent Hence, LO modes with $A_u$ symmetry are polarized along axis $\mathbf{b}$ only. LO modes with $B_u$ symmetry are polarized within the $\mathbf{a}-\mathbf{c}$ plane. The eigen displacement vectors, $\mathbf{\hat{e}}_{\stext{LO}_{(n)}}=\cos\alpha_{\stext{LO}_{(n)}} \mathbf{x}+\sin\alpha_{\stext{LO}_{(n)}}\mathbf{y}$, can be found from 

\begin{equation} \label{eq:acLOalpha}
\tan\alpha_{\stext{LO}_{(n)}}=-\frac{\varepsilon_{xx}(\omega_{\stext{LO}_{(n)}})}{\varepsilon_{xy}(\omega_{\stext{LO}_{(n)}})}=-\frac{\varepsilon_{xy}(\omega_{\stext{LO}_{(n)}})}{\varepsilon_{yy}(\omega_{\stext{LO}_{(n)}})}.
\end{equation}

\noindent For $\beta$-Ga$_2$O$_3$, in the absence of free charge carrier contributions, 4 LO modes with $A_u$ symmetry and 8 LO modes with $B_u$ symmetry are obtained from Eq.~(\ref{eq:acLO}) and Eq.~(\ref{eq:bLO}), respectively.

\paragraph{LPP mode determination} For $\beta$-Ga$_2$O$_3$, in the presence of free charge carrier contributions, Eq.~(\ref{eq:eps:LO}) factorizes again into

\begin{equation} \label{eq:acLPP}
0=\varepsilon_{xx}(\omega_{\stext{LPP}_{(n)}})\varepsilon_{yy}(\omega_{\stext{LPP}_{(n)}})-\varepsilon^2_{xy}(\omega_{\stext{LPP}_{(n)}}).
\end{equation}

\noindent and

\begin{equation} \label{eq:bLPP}
0=\varepsilon_{zz}(\omega_{\stext{LPP}_{(n)}}).
\end{equation}

\noindent Hence, LPP modes with $A_u$ symmetry are polarized along axis $\mathbf{b}$ only. LPP modes with $B_u$ symmetry are polarized within the $\mathbf{a}-\mathbf{c}$ plane. The eigen displacement vectors, $\mathbf{\hat{e}}_{\stext{LPP}_{(n)}}=\cos\alpha_{\stext{LPP}_{(n)}} \mathbf{x}+\sin\alpha_{\stext{LPP}_{(n)}}\mathbf{y}$, can be found from 

\begin{equation} \label{eq:acLPPalpha}
\tan\alpha_{\stext{LPP}_{(n)}}=-\frac{\varepsilon_{xx}(\omega_{\stext{LPP}_{(n)}})}{\varepsilon_{xy}(\omega_{\stext{LPP}_{(n)}})}=-\frac{\varepsilon_{xy}(\omega_{\stext{LPP}_{(n)}})}{\varepsilon_{yy}(\omega_{\stext{LPP}_{(n)}})}.
\end{equation}

\noindent The presence of a free charge carrier plasma within $\beta$-Ga$_2$O$_3$ results in 5 LPP modes with $A_u$ symmetry and 12 LPP modes with $B_u$ symmetry, and which are obtained from Eq.~(\ref{eq:acLPP}) and Eq.~(\ref{eq:bLPP}), respectively.

\paragraph{Lyddane-Sachs-Teller relation:} In the absences of free charge carriers, static and high frequency dielectric constants fulfill the Lyddane-Sachs-Teller (LST) relation~\cite{Lyddane41,Cochran62,Kittel2009}

\begin{equation}\label{eq:LST}
\frac{\varepsilon_{\stext{DC}}}{\varepsilon_{\infty}}=\prod^{m}_{l=1}\left(\frac{\omega_{\stext{LO},l}}{\omega_{\stext{TO},l}}\right)^2,
\end{equation}  

\noindent where $m$ denotes the number of mode branches of a given material along a given major polarizability axis. The LST relation is derived from the behavior of a dielectric function at static and high frequencies where the imaginary part must vanish. Because the long wavelength dielectric function can typically be rendered as a general response function with second order poles and zeros, the summation of all zeros and poles at static frequency leads to Eq.~(\ref{eq:LST}). Written most commonly with the intent for isotropic materials, the relation has been found correct for anisotropic dielectrics with orthogonal axes~\cite{SchubertPRB61_2000,SchoecheJAP2013TiO2,SchubertIRSEBook_2004}. It is also valid for the $\mathbf{b}$-axis response, i.e., for $\varepsilon_{zz}$ here.  

For the $\mathbf{a}$-$\mathbf{c}$ plane a physically meaningful set of dielectric functions along fixed orthogonal axes does not exist, and the relation in Eq.~(\ref{eq:LST}) is not generally valid for materials with monoclinic and triclinic crystal structures. However, a generalized relation for monoclinic materials can be found, analogous to the LST relation. Following the same logic in derivation, one may inspect the behavior of the sub determinant of the monoclinic dielectric function tensor, $\varepsilon_{xx}$$\varepsilon_{yy}$-$\varepsilon^2_{xy}$. At zero frequencies, this function is equal to $\varepsilon_{\stext{DC},xx}$$\varepsilon_{\stext{DC},yy}$-$\varepsilon^2_{\stext{DC},xy}$, the high frequency limit follows likewise. Casting the sub determinant into a factorized form, it is crucial to recognize that all terms with $(\omega^2_{\stext{TO},(l)}-\omega^2)^{-2}$ do not contribute to the summation because their amplitudes cancel. Hence, the denominator factorizes into the second order poles at all $B_u$ TO frequencies, and the numerator factorizes into all roots of the sub determinant. The order of the polynomials are both $2m$, hence, there are $m$ poles at $\omega^2_{\stext{TO},(l)}$ and $m$ zeros at $\omega^2_{\stext{LO},(l)}$. The generalized LST relation for monoclinic materials reads then

\begin{equation}\label{eq:LSTxy}
\frac{\varepsilon_{\stext{DC},xx}\varepsilon_{\stext{DC},yy}-\varepsilon^2_{\stext{DC},xy}}{\varepsilon_{\infty,xx}\varepsilon_{\infty,yy}-\varepsilon^2_{\infty,xy}}=\prod^{m}_{l=1}\left(\frac{\omega_{\stext{LO},l}}{\omega_{\stext{TO},l}}\right)^2.
\end{equation}   

In the above equation, $m$=8 denotes the number of modes with $B_u$ symmetry for $\beta$-Ga$_2$O$_3$. While the implementation of the LST relation, or its generalization above, is not truly needed when analyzing long wavelength ellipsometry data, the relations are quite useful to check for consistency of determined phonon and dielectric constant parameters.

\subsection{Generalized Ellipsometry}

For optically anisotropic materials it is necessary to apply the generalized ellipsometry approach because coupling between the $p$ (parallel to the plane of incidence) and $s$ (perpendicular to the plane of incidence) polarized incident electromagnetic plane wave components occurs upon reflection off the sample surface. $\beta$-Ga$_2$O$_3$ possesses monoclinic crystal structure, and is highly anisotropic. In previous work, which included uniaxial and biaxial materials in single layer and multiple layer structures such as corundum~\cite{SchubertPRB61_2000}, rutile~\cite{SchoecheJAP2013TiO2}, antimonite~\cite{Schubert03g}, pentacene~\cite{DresselOE2008pentacene,Schubert03c}, zinc metal oxides~\cite{Ashkenov03}, wurtzite structure group-III Nitride heterostructures~\cite{KasicPRB62_2000,Kasic02,DarakchievaAPL84_2004,DarakchievaPRB2004AlN,DarakchievaPRB2005AlGaNSL,DarakchievaPRB2007StrainGaN,DarakchievaAPL2009InNe,DarakchievaAPL2009InN,DarakchievaAPL2010InN,DarakchievaPRB2014InNmix,DarakchievaJAP2014InNMg}, and form induced anisotropic thin films~\cite{HofmannTHzGLADChapter2013} we discussed theory and applications of generalized ellipsometry in detail. In a number of recent publications we discussed treatment and necessity of investigating off axis cut surfaces from anisotropic crystals to gain access to all long wavelength active phonon modes, for example  in ZnO~\cite{Bundesmann03b}, and in wurtzite structure group-III Nitrides~\cite{Darakchievapssb2006aGaN,DarakchievaJCG2007acGaN,DarakchievaChapter2008}. A multiple sample, multiple azimuth, and multiple angle of incidence approach is required for $\beta$-Ga$_2$O$_3$. Hence, multiple single crystalline samples cut under different angles from the same crystal must be investigated and analyzed simultaneously. 

In the generalized ellipsometry formalism, the interaction of electromagnetic plane waves with layered samples is described within the Jones or Mueller matrix formalism~\cite{SchubertIRSEBook_2004,Thompkins_2004,HumlicekHOE,Azzam95}. The Mueller matrix renders the optical sample properties at a given angle of incidence and sample azimuth, and data measured must be analyzed through a best match model calculation procedure. The sample azimuth is defined by a certain in plane rotation with respect to the laboratory coordinate system's $z$ axis, set by a given orientation within a sample surface ($x$-$y$ plane) and the plane of incidence ($x$-$z$ plane), where the $z$ axis is parallel to the sample normal, and the coordinate origin is at the sample surface. The sample azimuth angle is defined separately for each sample investigated here. 

In the generalized ellipsometry situation the Stokes vector formalism, where real-valued matrix elements connect the Stokes parameters of the electromagnetic plane waves before and after sample interaction, is an appropriate choice for casting the ellipsometric measurement parameters. The Stokes vector components are defined by $S_{0}=I_{p}+I_{s}$, $S_{1}=I_{p} - I_{s}$, $S_{2}=I_{45}-I_{ -45}$, $S_{3}=I_{\sigma + }-I_{\sigma - }$, where $I_{p}$, $I_{s}$, $ I_{45}$, $I_{-45}$, $I_{\sigma + }$, and $I_{\sigma - }$denote the intensities for the $p$-, $s$-, +45$^{\circ}$, -45$^{\circ}$, right handed, and left handed circularly polarized light components, respectively~\cite{Fujiwara_2007}. The Mueller matrix is defined by arranging incident and exiting Stokes vector into matrix form
\begin{equation}
\left( {{\begin{array}{*{20}c}
 {S_{0} } \hfill \\ {S_{1} } \hfill \\  {S_{2} } \hfill \\  {S_{3} } \hfill \\
\end{array} }} \right)_{\mathrm{output}} =
\left( {{\begin{array}{*{20}c}
 {M_{11} } \hfill & {M_{12} } \hfill \ {M_{13} } \hfill & {M_{14} } \hfill \\
 {M_{21} } \hfill & {M_{22} } \hfill \ {M_{23} } \hfill & {M_{24} } \hfill \\
 {M_{31} } \hfill & {M_{32} } \hfill \ {M_{33} } \hfill & {M_{34} } \hfill \\
 {M_{41} } \hfill & {M_{42} } \hfill \ {M_{43} } \hfill & {M_{44} } \hfill \\
\end{array} }} \right)
\left( {{\begin{array}{*{20}c}
 {S_{0} } \hfill \\ {S_{1} } \hfill \\  {S_{2} } \hfill \\  {S_{3} } \hfill \\
\end{array} }} \right)_{\mathrm{input}}.
\end{equation}

\subsubsection{Ellipsometry data and model dielectric function analyses}

\begin{figure}[!tbp]
  \begin{center}
    \includegraphics[width=0.4\linewidth]{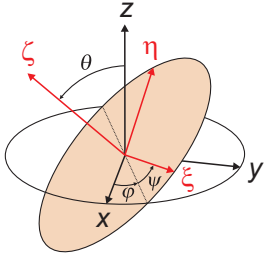}
    \caption{Definition of the Euler angles $\varphi$, $\theta$, and $\psi$ and the orthogonal rotations as provided by $\mathbf{A}$. ($\xi, \eta, \zeta $), and ($x, y, z$) refer to the Cartesian auxiliary and laboratory coordinate systems, respectively. Redrawn from Ref.~\cite{SchubertIRSEBook_2004}.}
    \label{fig:Euler}
  \end{center}
\end{figure}

Spectroscopic ellipsometry is an indirect method and requires detailed model analysis procedures in order to extract relevant physical parameters~\cite{JellisonHOE_2004,Aspnes98}. Here, the simple two phase (substrate ambient) model is employed, where the substrate represents single crystal $\beta$-Ga$_2$O$_3$ samples. The light propagation within the anisotropic substrate is calculated by applying a 4$\times$4 matrix algorithm applicable to plane parallel interfaces~\cite{Schubert96,Schubert03h,Schubert04}. The matrix algorithm requires a full description of all dielectric function tensor elements of the substrate. The full dielectric tensor is obtained by setting $\varepsilon_{xx}, \varepsilon_{xy}, \varepsilon_{yy}$, and $\varepsilon_{zz}$ as unknown parameters, and by setting the remaining elements to zero. Then, according to the crystallographic surface orientation of a given sample, and according to its azimuth orientation relative to the plane of incidence, a Euler angle rotation is applied to $\varepsilon$. The definition of the Euler angle parameters relative to the coordinate system used here is shown in Fig.~\ref{fig:Euler}. The Euler rotation parameters describe the angular positions of the auxiliary Cartesian system (Fig.~\ref{fig:Ga2O3unitcell}), here fixed by choice to the unit cell axes of $\beta$-Ga$_2$O$_3$, relative to the laboratory coordinate system for every ellipsometry measurement. Matrix A is obtained by

\begin{equation}
A=R_1(\varphi)R_2(\theta)R_1(\psi),
\end{equation}

\noindent with 

\begin{equation}
R_1(v)=
\left(\begin{array}{ccc}
\cos v & -\sin v & 0 \\
\sin v &  \cos v & 0 \\
0 & 0 & 1
\end{array}\right),
\end{equation}
\begin{equation}
R_2(v)=
\left(\begin{array}{ccc}
1 & 0 & 0 \\
0 & \cos v & -\sin v \\
0 & \sin v &  \cos v
\end{array}\right).
\end{equation}

As first step in data analysis, all ellipsometry data were analyzed using a wavelength by wavelength approach. Thereby, all data obtained at the same wavenumber from multiple samples, multiple azimuth angles, and multiple angles of incidence are included (polyfit) and one set of complex values $\varepsilon_{xx}, \varepsilon_{xy}, \varepsilon_{yy}$, and $\varepsilon_{zz}$ is searched for. This procedure is simultaneously performed for all wavelengths, while results of $\varepsilon_{xx}, \varepsilon_{xy}, \varepsilon_{yy}$, and $\varepsilon_{zz}$ for one wavelength have no influence on results at any other wavelength. In addition, each sample requires one set of 3 independent Euler angle parameters. The latter describe the rotations of the $\beta$-Ga$_2$O$_3$ auxiliary coordinate system at zero azimuth. Zero azimuth is the first azimuth position at which measurements were performed. Multiple azimuth positions differ by 45$^{\circ}$ counterclockwise increments. These increments are added to Euler angle parameter $\varphi$, and hence once the zero azimuth position parameter is known all other Euler parameters are known. In this polyfit and wavelength by wavelength approach, we have not augmented any physical lineshape assumptions for the spectral behavior of $\varepsilon_{xx}, \varepsilon_{xy}, \varepsilon_{yy}$, and $\varepsilon_{zz}$. In a second step, $\varepsilon_{xx}, \varepsilon_{xy}, \varepsilon_{yy}$, and $\varepsilon_{zz}$ are analyzed simultaneously by Eqs.~(\ref{eq:epsmonoall}). As a result, we obtain all parameters for TO, LO, and LPP modes as well as for static and high frequency dielectric constants.  

Two regression analyses (Levenberg-Marquardt algorithm) are performed. The first is minimizing the difference between measured and calculated generalized ellipsometry data during the polyfit. The second is minimizing the difference between the wavelength by wavelength extracted $\varepsilon_{xx}, \varepsilon_{xy}, \varepsilon_{yy}$, and $\varepsilon_{zz}$ spectra and those calculated by Eqs.~(\ref{eq:epsmonoall}). All model parameters were varied until calculated and experimental data matched as close as possible (best match model). This is done by minimizing the mean square error ($\chi^2$) function which is weighed to estimated experimental errors ($\sigma$) determined by the instrument for each data point~\cite{SchubertPRB61_2000,Schubert03h,SchubertIRSEBook_2004,SchubertADP15_2006,SchoecheJAP2013TiO2}. For the second regression step, the numerical uncertainty limits of the 90\% confidence interval from the first regression were used as experimental error bars for the wavelength by wavelength extracted $\varepsilon_{xx}, \varepsilon_{xy}, \varepsilon_{yy}$, and $\varepsilon_{zz}$ spectra. A similar approach was described, for example, in Refs.~\cite{SchubertPRB61_2000,HofmannPRB66_2002,SchubertIRSEBook_2004,SchoecheJAP2013TiO2}. All best match model calculations were performed using WVASE32 (J.~A.~Woollam~Co.,~Inc.)

\subsection{Phonon mode calculations}

\begin{table*}[!t]
\caption{Phonon mode parameters for $A_u$ and $B_u$ modes obtained from DFT calculations using Quantum Espresso. Renderings of displacements are shown in Fig.~\ref{fig:phononrendering}.}
\begin{center}
\begin{tabular}{lccccccccccccc}
    \noalign{\bigskip} \hline \hline
		& &$X=B_u$ & & & & & & & &$X=A_u$ & & & \\
    \cline{3-10}\cline{11-14}
    Parameter& &k=1&2&3&4&5&6&7&8&k=1&2&3&4\\
    \noalign{\smallskip} \hline
    $A_{k}^{X}$ & this work &4.65&9.48&30.57&28.1&5.37&0.89&7.33&10.43&12.76&23.24&14.34&0.07\\
    $\omega_{\stext{TO},k}$ [cm$^{-1}$] &this work& 753.76&705.78&589.86&446.83&365.84&289.71&260.4&202.4&678.39&475.69&327.45&155.69\\
		$\alpha_{\stext{TO},k}$ [$^{\circ}$]&this work &71&24&128&46&166&173&175&78& - & -& - & -\\ \hline 
		$\omega_{\stext{TO},k}$ [cm$^{-1}$] &Ref.~\cite{LiuAPL2007Ga2O3phononcalc} &741.6&672.6&574.3&410.5&343.6&265.3&251.6&187.5&647.9&383.5&296.2&141.6\\
		\hline \hline 
\end{tabular} \label{tab:TOAuBuQE}
\end{center}
\end{table*}

\begin{figure*}[!tbp]
  \begin{center}
         \includegraphics[width=1.\linewidth]{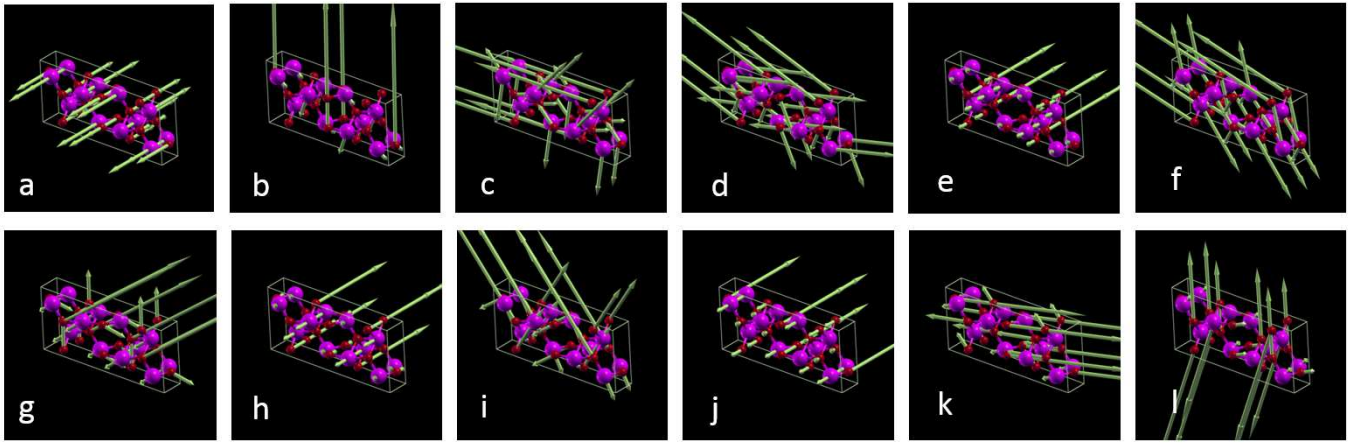}
   \caption{Renderings of TO phonon modes in $\beta$-Ga$_2$O$_3$ with $A_u$ (a: $A_u(4)$, e: $A_u(3)$, h: $A_u(2)$, j: $A_u(1)$) and $B_u$ symmetry (b: $B_u(8)$, c: $B_u(7)$, d: $B_u(6)$, f: $B_u(5)$, g: $B_u(4)$, i: $B_u(3)$, k: $B_u(2)$, l: $B_u(1)$). The respective phonon mode frequency parameters calculated using Quantum Espresso are given in Tab.~\ref{tab:TOAuBuQE}, and obtained using XCrysDen~\cite{KokaljCMS2003XCrysDen,XCrysDen}.}
    \label{fig:phononrendering}
  \end{center}
\end{figure*}

Theoretical calculations of long wavelength active $\Gamma$-point phonon frequencies were performed by plane wave density functional theory (DFT) using Quantum ESPRESSO (QE)~\cite{GiannozziJPCM2009QE}. The exchange correlation functional of Perdew and Zunger (PZ)~\cite{PerdewPRB1981} and norm conserving pseudopotentials from the QE library were implemented. A primitive cell of $\beta$-Ga$_2$O$_3$ consisting of six Oxygen and four Gallium atoms was first relaxed to force levels less than 1/1000 Ry/Bohr. A dense $4\times8\times16$ regular Monkhorst-Pack grid was used for sampling of the Brillouin Zone~\cite{MonkhorstPRBGRID}. A convergence threshold of $1\times10^{-12}$ was used to reach self consistency with a large electronic wavefunction cutoff of 100 Ry. The phonon frequencies were computed at the $\Gamma$-point of the Brillouin zone using density functional perturbation theory for phonons~\cite{BaroniRMP2001DFTPhonons}. Results for long wavelength active modes with $A_u$ and $B_u$ symmetry are listed in Tab.~\ref{tab:TOAuBuQE}. Data listed include the TO resonance frequencies, and the eigenvector angle relative to axis $\mathbf{c}$ within the $\mathbf{a}-\mathbf{c}$ plane for modes with $B_u$ symmetry. Renderings of molecular displacements for each mode are shown in Fig.~\ref{fig:phononrendering}.

\section{Experiment}

Single crystals of $\beta$-Ga$_2$O$_3$  were grown by the edge-defined film-fed growth method described in Refs.~\cite{AidaEDFFGa2O3,SasakiAPE2012,ShimamuraAPPAGa2O3} at Tamura Corp., Japan. The substrates were fabricated by slicing from bulk crystals according to their intended surface orientation, and then single side polished. The substrate dimensions are 650$\mu$m$\times$10$\times$10~mm$^2$. The substrates are Sn doped with an estimated activated electron density of $N_d-N_a\approx (2-9)\times 10^{18}$cm$^{-3}$.

The vibrational properties and free charge carrier properties of $\beta$-Ga$_2$O$_3$ were studied by room temperature infrared (IR) and farinfrared (FIR) GSE. The IR-GSE measurements were performed on a rotating compensator infrared ellipsometer (J.~A.~Woollam Co., Inc.) in the spectral range of 500 -- 1500 cm$^{-1}$ with a spectral resolution of 2 cm$^{-1}$. The FIR-GSE measurements were performed on a in-house built rotating polarizer rotating analyzer farinfrared ellipsometer in the spectral range of 50 -- 500 cm$^{-1}$ with an average spectral resolution of 1 cm$^{-1}$~\cite{KuehneRSI2014}. All GSE measurements were performed at 50$^\circ$, 60$^\circ$, and 70$^\circ$ angles of incidence. All measurements are reported in terms of Mueller matrix elements, which are normalized to element $M_{11}$. Note that due to the lack of a compensator for the FIR range in this work, no elements of fourth row or column is reported for the FIR range. In order to acquire sufficient information to differentiate and determine $\varepsilon_{xx}, \varepsilon_{xy}, \varepsilon_{yy}$, and $\varepsilon_{zz}$, data measured from at least two differently cut surfaces of $\beta$-Ga$_2$O$_3$, and within at least two different azimuth positions are needed. Here, we investigate a (010) and a ($\bar{2}$01) sample. At least 5 azimuth positions were measured on each sample, separated by 45$^{\circ}$.

\section{Results and discussion}

\subsection{Dielectric Function Tensor analysis}

\begin{figure*}[!tbp]
  \begin{center}
         \includegraphics[width=1\linewidth]{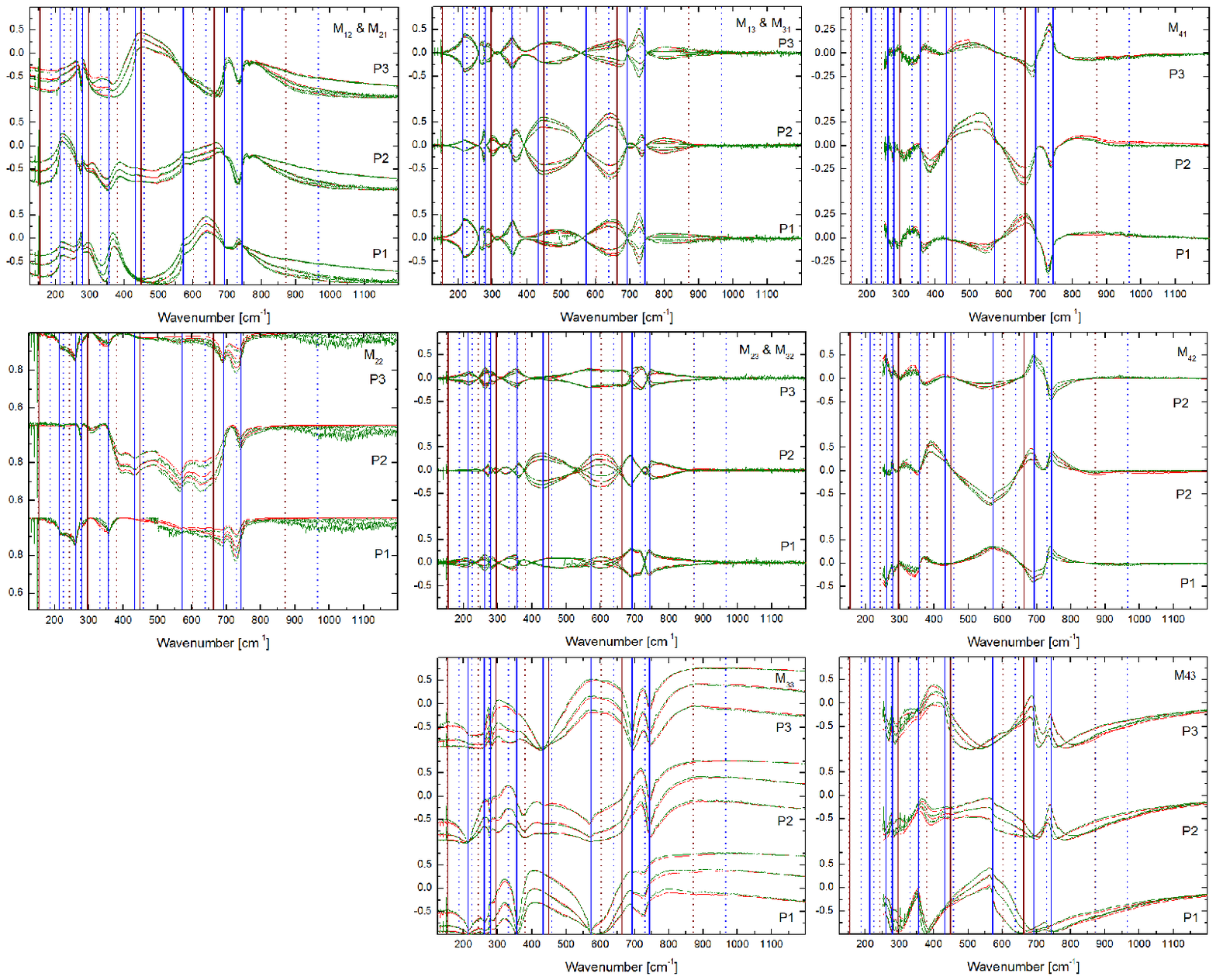}
   \caption{Experimental (dotted, green lines) and best match model calculated (solid, red lines) Mueller matrix data obtained from a (010) surface at three different sample azimuth orientations. (P1: $\varphi=62.5(4)^{\circ}$, P2: $\varphi=107.5(4)^{\circ}$, P3: $\varphi=152.5(4)^{\circ}$). Data were taken at three angles of incidence ($\Phi_a=50^{\circ}, 60^{\circ}, 70^{\circ}$). Equal Mueller matrix data, symmetric in their indices, are plotted within the same panels for convenience. Vertical lines indicate wavenumbers of TO (solid lines) and LPP modes (dotted lines) with $B_u$ symmetry (blue) and $A_u$ symmetry (brown). Fourth column elements are only available from the IR instrument. Note that all elements are normalized to $M_{11}$. The remaining Euler angle parameters are $\theta=0.4(2)$ and $\psi=0.0(1)$ consistent with the crystallographic orientation of the (010) surface. }
    \label{fig:exp010}
  \end{center}
\end{figure*}

\begin{figure*}[!tbp]
  \begin{center}
         \includegraphics[width=1\linewidth]{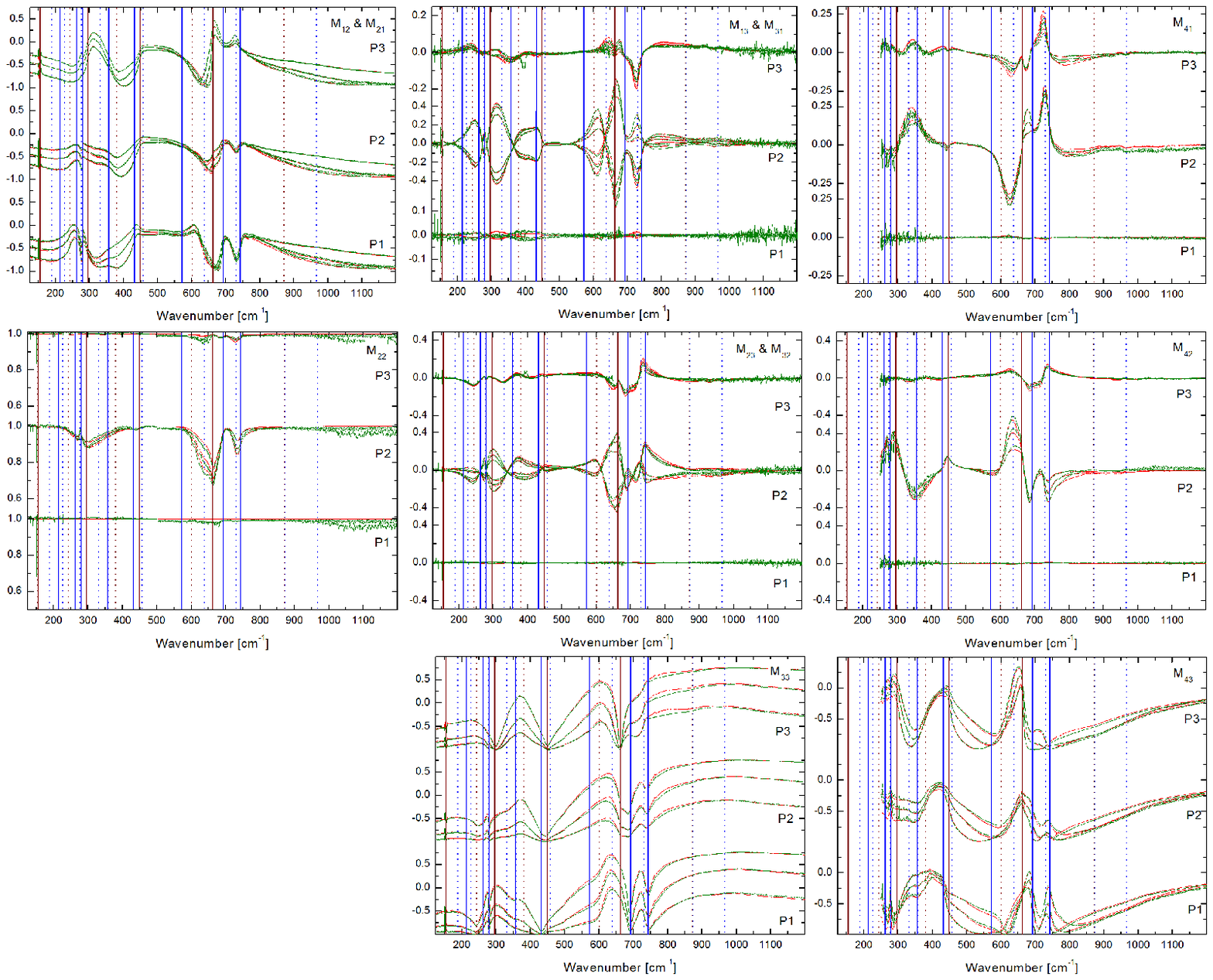}
   \caption{Same as Fig.~\ref{fig:exp010} for the ($\bar{2}$01) sample at azimuth orientation P1: $\varphi=179.(3)^{\circ}$, P2: $\varphi=224.(3)^{\circ}$, P3: $\varphi=269.(3)^{\circ}$. $\theta=90.(5)$ and $\psi=-28(1)$, consistent with the crystallographic orientation of the $(\bar{2}01)$ surface. }
    \label{fig:exp-201}
  \end{center}
\end{figure*}

Figures~\ref{fig:exp010} and~\ref{fig:exp-201} summarize experimental and best match model calculated data for the $(010)$ and $(\bar{2}01)$ surfaces investigated in this work. Graphs depict selected data, obtained at 3 different sample azimuth orientations each $45^{\circ}$ apart. Panels with individual Mueller matrix elements are shown separately, and individual panels are arranged according to the indices of the Mueller matrix element. It is observed by experiment as well as by model calculations that all Mueller matrix elements are symmetric, i.e., $M_{ij}=M_{ji}$.  
Hence, elements with $M_{ij}=M_{ji}$, i.e., from upper and lower diagonal parts of the Mueller matrix, are plotted within the same panels. Therefore, the panels represent the upper part of a $4\times4$ matrix arrangement. Because all data obtained are normalized to element $M_{11}$, and because $M_{1j}=M_{j1}$, the first column does not appear in this arrangement. The only missing element is $M_{44}$, which cannot be obtained in our current instrument configuration due to the lack of a second compensator. Data are shown for wavenumbers (frequencies) from 125 cm$^{-1}$ to 1200 cm$^{-1}$, except for column $M_{4j}=M_{j4}$ which only contains data from approximately 250 cm$^{-1}$ to 1200 cm$^{-1}$. All other panels show data obtained within the FIR range (125 cm$^{-1}$ to 500 cm$^{-1}$) using our FIR instrumentation and data obtained within the IR range (500 cm$^{-1}$ to 1200 cm$^{-1}$) using our IR instrumentation. Data from the additional azimuth orientations (at least 2) for each sample are not shown. 

\begin{figure}[!tbp]
  \begin{center}
    \includegraphics[width=0.9\linewidth]{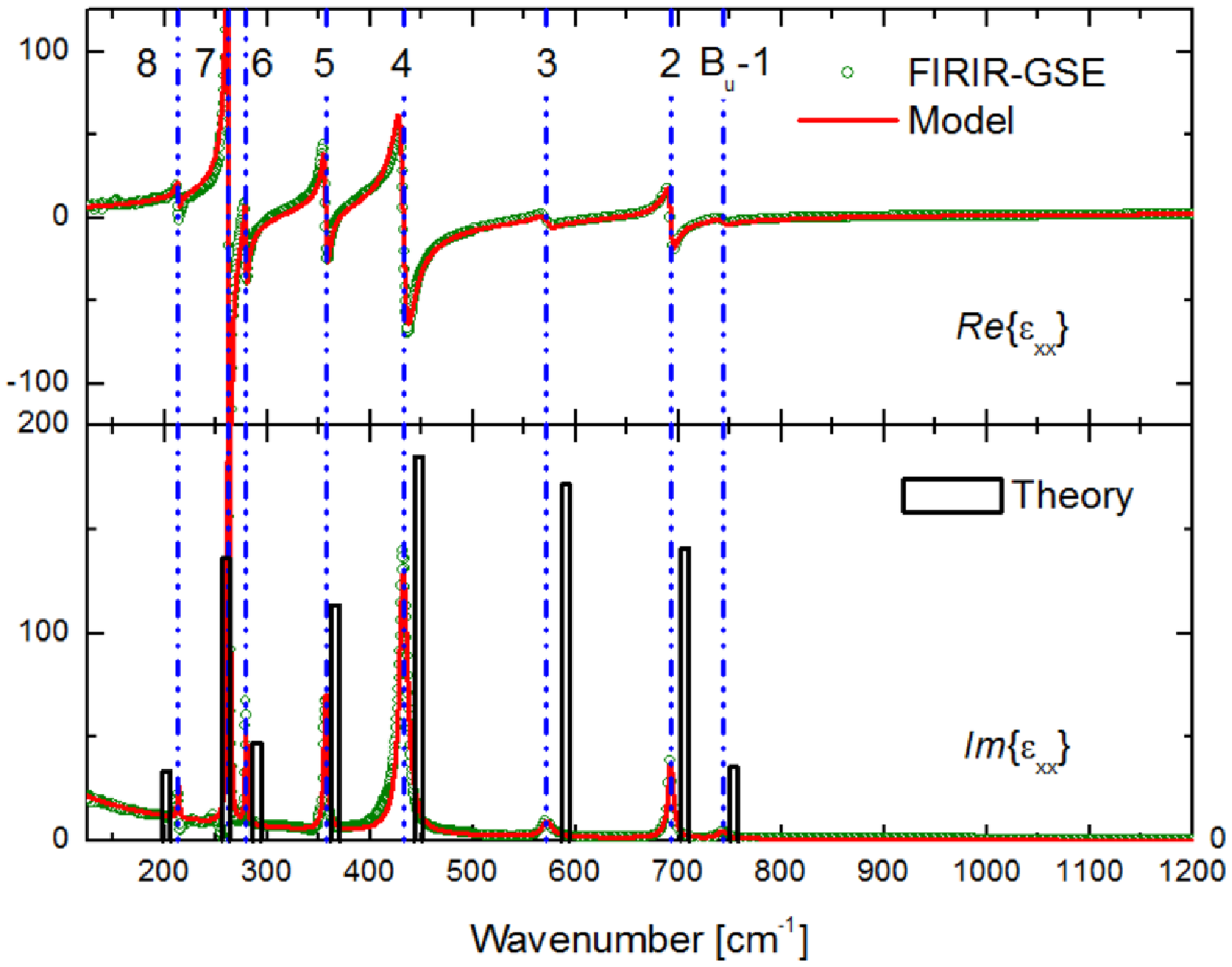}
    \caption{Dielectric function tensor element $\varepsilon_{xx}$, representative for axis $\mathbf{c}$. Lines indicate results from wavelength by wavelength best match model calculation  to experimental Mueller matix data (dotted; green) and best match model lineshape analysis (solid; red).Vertical lines indicate $B_u$ mode TO frequencies. Vertical bars indicate relative displacement amplitudes from phonon mode calculations projected onto direction $\mathbf{x}$.}
    \label{fig:epsxx}
  \end{center}
\end{figure}

\begin{figure}[!tbp]
  \begin{center}
    \includegraphics[width=.9\linewidth]{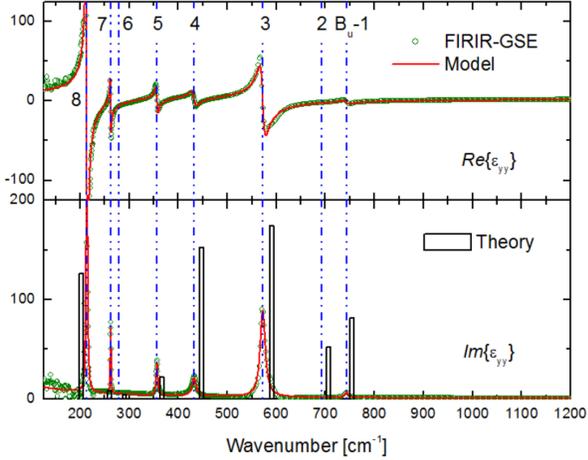}
    \caption{Same as Fig.~\ref{fig:epsxx} for $\varepsilon_{yy}$, representative for polarization along direction $\mathbf{y}$. The calculated displacement amplitudes are projected onto direction $\mathbf{y}$.}
    \label{fig:epsyy}
  \end{center}
\end{figure}

\begin{figure}[!tbp]
  \begin{center}
    \includegraphics[width=.9\linewidth]{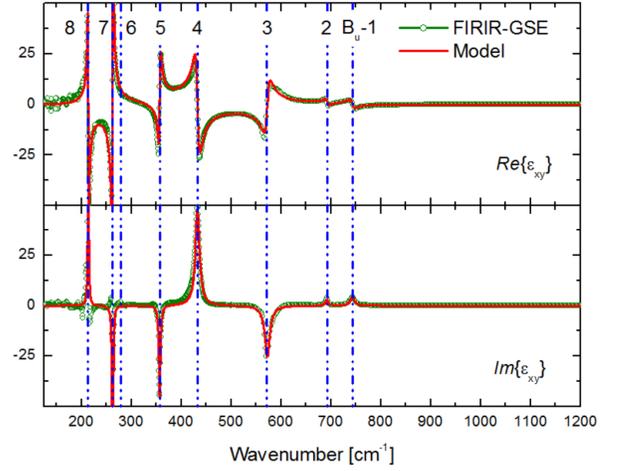}
    \caption{Same as Fig.~\ref{fig:epsxx} for $\varepsilon_{xy}$, the shear transformation element within the $\mathbf{a}$-$\mathbf{c}$ plane.}
    \label{fig:epsxy}
  \end{center}
\end{figure}

\begin{figure}[!tbp]
  \begin{center}
    \includegraphics[width=.9\linewidth]{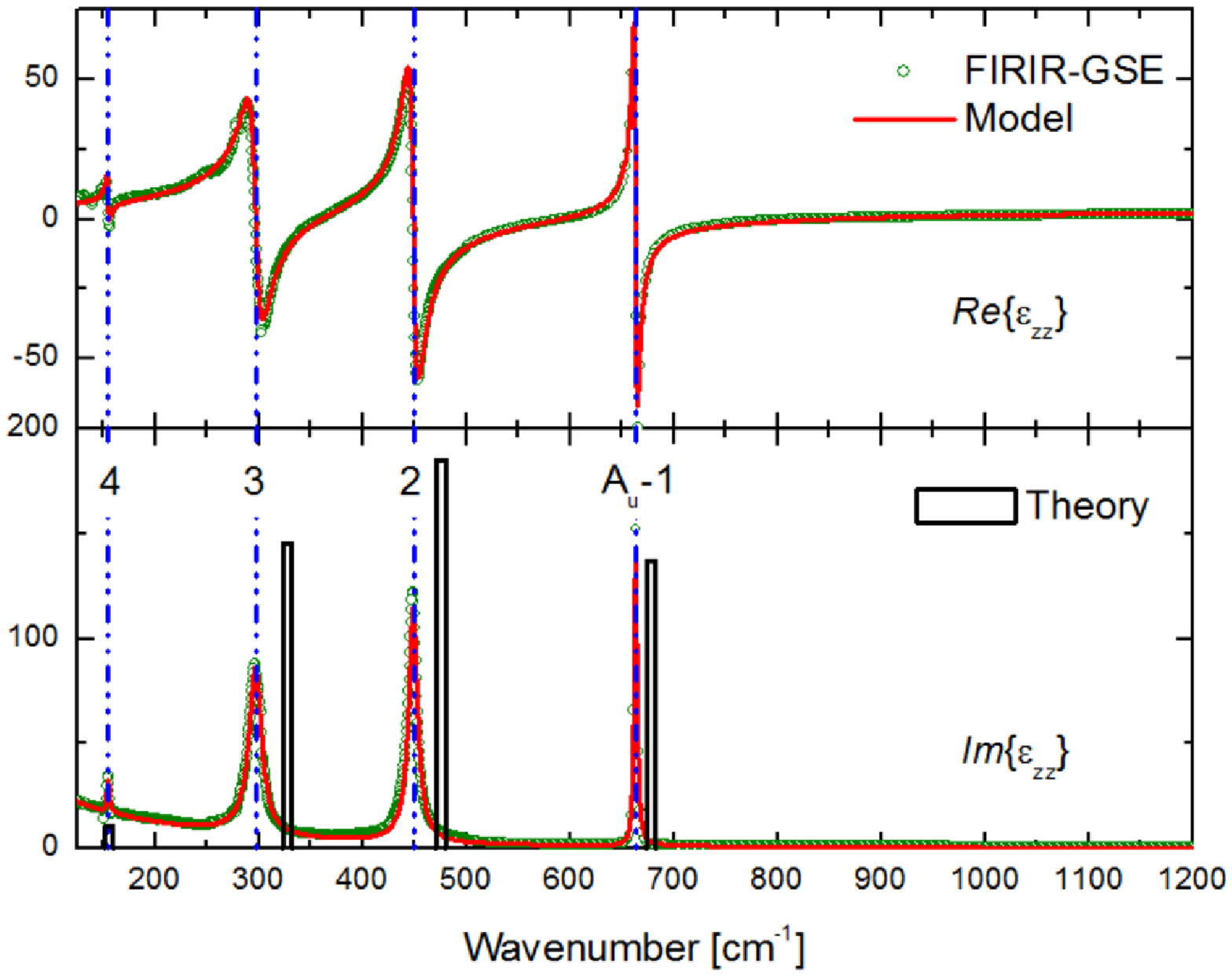}
    \caption{Same as Fig.~\ref{fig:epsxx} for $\varepsilon_{zz}$, representative for polarization along axis $\mathbf{b}$. Vertical lines indicate $A_u$ mode TO frequencies. The calculated displacement amplitudes are projected onto direction $\mathbf{z}$.}
    \label{fig:epszz}
  \end{center}
\end{figure}

While every data set (sample, position, azimuth, angle of incidence) is unique, all data sets share characteristic features at certain wavelengths. These wavelengths are indicated by vertical lines. As discussed further below, all lines are associated with TO or LPP modes with either $A_u$ or $B_u$ symmetry. While we do not show all data in Figures~\ref{fig:exp010} and~\ref{fig:exp-201} for brevity, we note that all data sets possess a twofold azimuth symmetry, i.e., all data sets are identical when a sample is measured again shifted by $180^{\circ}$ azimuth orientation. The most notable observation from the experimental Mueller matrix data behavior is the strong anisotropy which is reflected by the non vanishing off diagonal block elements $M_{13}$, $M_{23}$, $M_{14}$, and $M_{24}$, and the strong dependence on sample azimuth in all elements. All data were analyzed simultaneously during the polyfit, best match model data regression procedure. For every wavelength,  up to 330 independent data points were included from the different samples, azimuth positions, and angles of incidence, while only 8 independent parameters for $\varepsilon_{xx}$, $\varepsilon_{xy}$, $\varepsilon_{yy}$, and $\varepsilon_{zz}$ were searched for. In addition, two sets of 3 wavelength independent Euler angle parameters were looked for. The results of this calculation are shown in Figs.~\ref{fig:exp010} and~\ref{fig:exp-201} as solid lines for the Mueller matrix elements, and in Figs.~\ref{fig:epsxx},~\ref{fig:epsyy},~\ref{fig:epsxy}, and~\ref{fig:epszz} as dotted lines for $\varepsilon_{xx}$, $\varepsilon_{xy}$, $\varepsilon_{yy}$, and $\varepsilon_{zz}$, respectively. In Figures~\ref{fig:exp010} and~\ref{fig:exp-201} the agreement between measured and model calculated data is excellent. The Euler angle parameters, given in captions of Figs.~\ref{fig:exp010} and~\ref{fig:exp-201} are in excellent agreement with the orientations of the crystallographic sample axes. For example, measurement on sample (010) initiated with axis $\mathbf{b}$ parallel to direction $\mathbf{z}$, and a natural cleavage edge parallel to $\mathbf{c}$ was oriented approximately with $30^{\circ}$ azimuth from the plane of incidence. 

To begin with, distinct features in $\varepsilon_{xx}$, $\varepsilon_{xy}$, $\varepsilon_{yy}$, and $\varepsilon_{zz}$ can be discussed without further model lineshape calculations. Vertical lines are drawn in the figures to indicate extrema in the imaginary parts of each element. One can observe that these vertical lines are identical for $\varepsilon_{xx}$, $\varepsilon_{xy}$, and $\varepsilon_{yy}$, while a different set is seen for $\varepsilon_{zz}$. There are 8 distinct frequencies in $\varepsilon_{xx}$, $\varepsilon_{xy}$, and $\varepsilon_{yy}$, and 4 in $\varepsilon_{zz}$. These frequencies indicate TO modes with $B_u$ and $A_u$ symmetry. The vertical line indexed with $``6"$ in $\varepsilon_{xx}$, $\varepsilon_{xy}$, and $\varepsilon_{yy}$ is associated with a resonance feature which only occurs in $\varepsilon_{xx}$. This indicates a mode with polarization along direction $\mathbf{c}$ only, while all other lines indicate modes which are neither polarized purely along $\mathbf{x}$ nor $\mathbf{y}$. We further note the asymptotic increase towards longer wavelengths in the imaginary parts of $\varepsilon_{xx}$, $\varepsilon_{yy}$, and $\varepsilon_{zz}$. This increase is likely caused by free charge carrier contributions. No such behavior is seen in $\varepsilon_{xy}$. 

\subsection{Phonon mode analysis}

The imaginary parts of $\varepsilon_{xx}$, $\varepsilon_{yy}$, and $\varepsilon_{zz}$ show features, which can typically be rendered by the Lorentzian broadened harmonic oscillator functions in Eq.~(\ref{eq:xhiij}). With our model introduced in Sect.~\ref{sec:AnisDFTensor} we obtain best match model calculations, which are also shown in Figs.~\ref{fig:epsxx}-~\ref{fig:epszz}. Again, an excellent match between the wavelength by wavelength determined dielectric function tensor elements and our physical model lineshape rendering is noted. It is worthwhile noting that the wavelength by wavelength derived dielectric functions are all Kramers-Kronig consistent since the Lorentzian broadened harmonic oscillator functions are Kramers-Kronig consistent. We have thereby independently verified that all tensor components of $\beta$-Ga$_2$O$_3$ are Kramers-Kronig consistent. The best match model lineshape calculation parameters are summarized in Tab.~\ref{tab:TOAuBuGSE}. As a result, we obtain phonon mode parameters for TO, LO, and LPP modes.

\paragraph{TO and LO modes:} We find 8 TO mode frequencies within elements $\varepsilon_{xx}$, $\varepsilon_{xy}$, and $\varepsilon_{yy}$. These are the modes with $B_u$ symmetry. The vertical lines and mode indices in Figs.~\ref{fig:epsxx},~\ref{fig:epsyy}, and~\ref{fig:epsxy} are located at frequencies which are identical with frequencies for $\omega_{\stext{TO}}$ listed in Tab.~\ref{tab:TOAuBuGSE}. As discussed in Sect.~\ref{sec:DFTensorModel}, element $\varepsilon_{xy}$ provides insight into the relative orientation of the unit eigen displacement vectors for each TO mode within the $\mathbf{a}$-$\mathbf{c}$ plane. In particular, modes $B_u$-3, $B_u$-5, and $B_u$-7 cause negative imaginary resonance features in $\varepsilon_{xy}$. Accordingly, their unit eigen displacement vectors in Tab.~\ref{tab:TOAuBuGSE} reflect values larger than 90$^{\circ}$. Modes $B_u$-1, $B_u$-2, $B_u$-4, and $B_u$-8 possess values less than 90$^{\circ}$. Accordingly, their resonance features in the imaginary part of $\varepsilon_{xy}$ are positive. Mode $B_u$-6 does not produce a resonance feature in the imaginary part of $\varepsilon_{xy}$, and its unit eigen displacement vector is parallel to $\mathbf{c}$. Accordingly, as predicted, such a mode only produces features in $\varepsilon_{xx}$ and none in $\varepsilon_{yy}$. This is verified by our experimental finding here. 

The TO mode frequencies and their unit eigen displacement vectors obtained from the ellipsometry model analysis are in very good agreement with the DFT phonon mode calculations shown in Tab.~\ref{tab:TOAuBuQE}. Predicted mode frequencies agree within few wavenumbers with the experimental findings. Calculated angles $\alpha$ agree within less than 25$^{\circ}$ of those found from our model analysis of the dielectric function tensor elements. In further agreement, modes $B_u$-3, $B_u$-5, and $B_u$-7 are predicted by theory to show the experimentally observed angular values larger than 90$^{\circ}$, and modes $B_u$-1, $B_u$-2, $B_u$-4, and $B_u$-8 reveal by experiment the predicted angular values less than 90$^{\circ}$. Mode $B_u$-6, which we find parallel to $\mathbf{c}$ has a value predicted near 180$^{\circ}$, in agreement with our experimental finding. Note that the eigen displacement vectors describe a uni-polar property without a directional assignment. Hence, $\alpha$ and $\alpha\pm\pi$ render equivalent eigen displacement orientations.

\begin{figure}[!tbp]
  \begin{center}
    \includegraphics[width=1\linewidth]{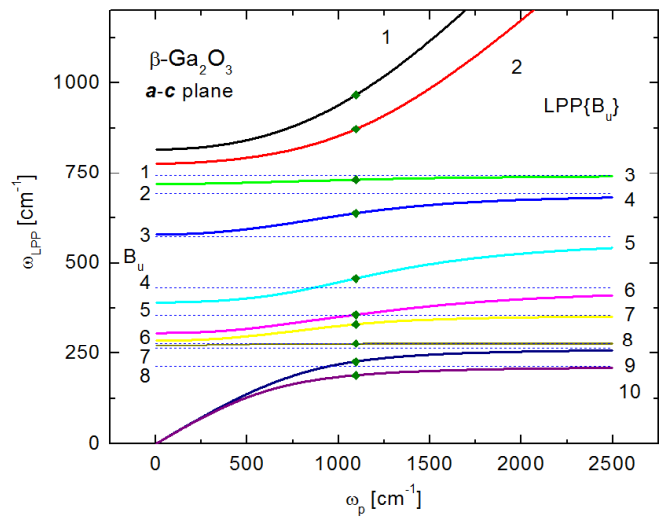}
    \caption{LPP coupled modes polarized within the $\mathbf{a}-\mathbf{c}$ plane as a function of isotropic plasma frequency $\omega_p$. The horizontal lines indicate the frequencies of the $B_u$ symmetry TO modes. Observed here is the deviation from the so called TO-LO-rule usually observed in semiconductor materials with orthogonal eigenpolarization systems, which is no longer valid for monoclinic lattices. Symbols (diamonds) indicate the LPP mode frequencies observed in FIR-GSE and IR-GSE spectra in this work. Numbering of modes as shown in Tab.~\ref{tab:LPPAuBuGSE}. Note that dispersion of LPP mode 8 is very small and within $B_u$ $\omega_{\stext{TO},6}$ and $\omega_{\stext{TO},7}$.}
    \label{fig:BuLPP}
  \end{center}
\end{figure}

Using Eq.~(\ref{eq:eps:LO}) one can calculate the intrinsic LO modes, that is, the LPP modes in the absence of free charge carriers. The free charge carrier properties are discussed further below. Subtracting the effects of the free charge carriers from the model functions for $\varepsilon_{xx}$, $\varepsilon_{xy}$, $\varepsilon_{yy}$, and $\varepsilon_{zz}$ the LO modes with $B_u$ and $A_u$ symmetry follow from Eqs.~(\ref{eq:acLO}) and~(\ref{eq:bLO}), respectively. We find 4 LO modes with $A_u$ and 8 LO modes with $B_u$ symmetry. Their values are summarized in Tab.~\ref{tab:TOAuBuGSE}. $B_u$ symmetry modes are also indicated in Fig.~\ref{fig:BuLPP} at $\omega_p=0$. 

\begin{figure}[!tbp]
  \begin{center}
    \includegraphics[width=1\linewidth]{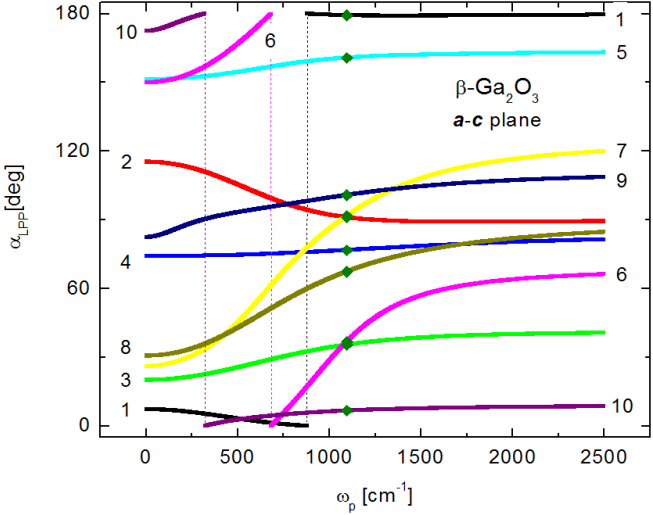}
    \caption{Unit eigen displacement vectors of the LPP coupled modes polarized within the $\mathbf{a}-\mathbf{c}$ plane as a function of isotropic plasma frequency $\omega_p$. Note that the free plasma like modes 1 and 2 approach the $x$ and $y$ axes for  $\omega_p \rightarrow \infty$. Symbols (diamonds) indicate vectors derived for the samples studied in this work. Numbering of modes as shown in Tab.~\ref{tab:LPPAuBuGSE}.}
    \label{fig:BuLPPvec}
  \end{center}
\end{figure}

In materials with multiple phonon modes, typically the TO-LO rule holds, i.e., a TO mode is always followed by an LO mode with increasing frequency (wavenumber). We note that the TO-LO rule is fulfilled for modes with $A_u$ symmetry, but not for $B_u$ symmetry (Fig.~\ref{fig:BuLPP}). This observation can be understood by inspecting the unit eigen displacement vectors. These are all parallel for TO and LO modes with $A_u$ symmetry. Hence, the displacement pattern at which the net displacement charge sum is zero (LO mode) occurs above a TO frequency, and is bound by the next TO frequency. The TO-LO splitting only depends on the polarity of the TO resonance. The polarity expresses itself as the amplitude of the TO resonance. At any TO resonance, the net displacement charge is non zero, and changes from positive to negative when moving across the TO frequency. Because the displacement pattern are disjunct between TO and LO modes, a LO mode cannot move across a TO mode, for example when the amplitudes of TO modes change. On the contrary, each TO and LO mode has a different orientation for modes with $B_u$ symmetry. In crystals with monoclinic symmetry, the TO-LO pattern distribution is 2-dimensional. The LO mode charge oscillations do not necessarily share the same direction with the TO oscillations. Hence, a LO mode pattern may form at a frequency which is larger than those from a pair of TO modes, if the TO modes each have different angles with each other as well as with the LO oscillation. The vectors for the LO modes with $B_u$ symmetry are shown in Fig.~\ref{fig:BuLPPvec} at $\omega_p=0$.

\begin{table*}[!t]
\caption{TO and LO phonon parameters for $A_u$ and $B_u$ modes obtained from best match model analysis of $\varepsilon_{xx}$, $\varepsilon_{xy}$, $\varepsilon_{yy}$, and $\varepsilon_{zz}$. Also shown are the eigenvector polarization angles for $B_u$ LO modes. The last digit which is determined within the 90\% confidence interval is indicated with brackets. Also included are data from recent IR reflectance measurements and phonon mode calculations. Data in square brackets were deduced assuming isotropic reflectance likely leading to erroneous TO parameters.}
\begin{center}
\begin{tabular}{lcccccccccccc}
    \noalign{\bigskip} \hline \hline
		&$X=B_u$ & & & & & & & &$X=A_u$ & & &\\
    \cline{2-9}\cline{10-13}
    Parameter&k=1&2&3&4&5&6&7&8&k=1&2&3&4\\
    \noalign{\smallskip} \hline
    $A_{k}^{X}$ [cm$^{-2}$] & 256.4(5)& 426.8(7)&820.3(6)&792.8(5)&358.3(7)&161.(7)&485.6(7)&520.7(5)&542.(5)& 718.(8) &579.(8)&7(2).\\
    $\omega_{\stext{TO},k}$ [cm$^{-1}$] &743.5(5)&692.4(4)&572.5(3)&432.5(6)&356.8(1)&279.1(5)&262.3(8)&213.7(9)& 663.2(2)& 448.6(5)&296.6(4)&154.8(5)\\
		$\gamma_{\stext{TO},k}$ [cm$^{-1}$]& 10.(4)& 6.4(4)&12.3(2)&10.0(5)&3.7(9)&1.8(5)&1.7(5)&1.9(8)&3.2(3)&10.2(8)&14.3(1)&2.(1)\\
		$\alpha_{\stext{TO},k}$ [$^{\circ}$]&48.(7)&5.(4)&10(6).&21.(0)&14(4).&0.(0)&158.(5)&80.(9)&-&-& - & -\\
		$\omega_{\stext{LO},k}$ [cm$^{-1}$]&817.(0)&778.(1)&719.(1)&579.(3)&391.(8)&307.(5)&286.(5)&271.(2)&770.(3)&558.(9)&344.(7)&156.(0)\\
		$\alpha_{\stext{LO},k}$ [deg]&7.(2)&11(5).&20.(1)&74.(2)&15(1).&15(0).&26.(1)&30.(7)& - & - & - & -\\
		\hline
		$\omega_{\stext{TO},k}$ [cm$^{-1}$]&[779]$^a$&[737]$^a$&[631]$^a$&[537]$^a$&[372]$^a$&[298]$^a$&[276]$^a$&[223]$^a$&660$^b$&449$^b$&295$^b$&$\approx$220$^b$\\
		$\omega_{\stext{LO},k}$ [cm$^{-1}$]&746.6$^c$&728.2$^c$&625.3$^c$&484.7$^c$&354.1$^c$&283.6$^c$&264.5$^c$&190.5$^c$&738.5$^c$&510.6$^c$&325.5$^c$&146.5$^c$\\
		\hline \hline 
		& \multicolumn{5}{c}{$^a$IR Refl.$||$ $\mathbf{c}$, Ref.~\cite{Villorapssa2002Ga2O3IR}.}\\ 
		& \multicolumn{5}{c}{$^b$IR Refl.$||$ $\mathbf{b}$, Ref.~\cite{Villorapssa2002Ga2O3IR}.}\\ 
		& \multicolumn{5}{c}{$^c$Theory, Ref.~\cite{LiuAPL2007Ga2O3phononcalc}.}\\ 
\end{tabular} \label{tab:TOAuBuGSE}
\end{center}
\end{table*}

Included into Figs.~\ref{fig:epsxx},~\ref{fig:epsyy}, and~\ref{fig:epszz} are the magnitudes of the calculated phonon displacements using Quantum Epsresso by vertical bars. The bars are located at the calculated frequencies of the TO modes. Overall, we note a good agreement between calculated and experimental results within few wavenumbers. The magnitude of the absorption features within the imaginary parts of the dielectric function tensor elements are comparable with the calculated displacement amplitudes. The agreement for $\varepsilon_{zz}$ is very good, and quite reasonable for the $B_u$ modes, except for mode $B_u$-3 which appears to overestimate the contribution to $\varepsilon_{xx}$. 

Liu, Mu, and Liu studied the lattice dynamical properties of $\beta$-Ga$_2$O$_3$ by using density functional perturbation theory~\cite{LiuAPL2007Ga2O3phononcalc}. The TO modes are included in Tab.~\ref{tab:TOAuBuQE} for comparison with our theoretical calculation results. The modes agree reasonably well, except for $A_u$-2 (See Tab.~\ref{tab:TOAuBuGSE}.). However, for the latter mode our theoretical results are much closer to our experimental results than the theoretical calculation in Ref.~\cite{LiuAPL2007Ga2O3phononcalc}. We further included calculated LO modes from Ref.~\cite{LiuAPL2007Ga2O3phononcalc} in Tab.~\ref{tab:TOAuBuGSE}, however, we find their values are not in agreement at all with our experimental findings.

V\'{i}llora \textit{et al.} investigated single crystals $\beta$-Ga$_2$O$_3$ grown by the floating zone technique~\cite{Villorapssa2002Ga2O3IR}. Polarized reflectance spectra with an incidence angle of about 10$^{\circ}$ and in the $50-1200$ cm$^{-1}$ spectral region revealed 12 long wavelength active modes, and contributions due to free charge carriers. The authors reported TO mode parameters and plasma parameters, and compared with measurements of the electrical conductivity and the electrical Hall coefficient. Platelet samples with surface (100) orientation allowed reflectance measurements with polarization along axes $\mathbf{b}$ and $\mathbf{c}$. Not all modes could be resolved in all samples, and uncertainty limits were not provided. The TO mode frequencies obtained in our present work agree excellently with modes reported for $A_u$ symmetry in Ref.~\cite{Villorapssa2002Ga2O3IR}. However, the TO mode frequencies for $B_u$ symmetry reported by V\'{i}llora \textit{et al.} deviate substantially from those found in this present work (Tab.~\ref{tab:TOAuBuGSE}). We explain this substantial difference by the fact that the authors ignored the anisotropy in the monoclinic $\beta$-Ga$_2$O$_3$ samples. Instead, the authors assumed that the measured reflectance spectra for polarization along axes $\mathbf{b}$ and $\mathbf{c}$ can be analyzed individually by using isotropic Fresnel equations for model calculations. While this assumption is correct for polarization parallel to axis $\mathbf{b}$ (but valid at normal incidence only), it is incorrect for polarization along $\mathbf{c}$ regardless of the angle of incidence. For the latter case, the isotropic model cannot correctly account for contributions that originate from $\varepsilon_{xy}$. As a result, incorrect, virtual resonance features appear when matching Lorentzian lineshapes to the measured reflectance data. We strongly believe that this explains the substantial deviations between the modes reported by V\'{i}llora \textit{et al.} and the modes reported in this work. Bermudez and Prokes investigated $\beta$-Ga$_2$O$_3$ nanoribbons by infrared reflectance spectroscopy~\cite{BermudezLangmuir2007} but no quantitative model analysis of the reflectance spectra were provided.

\paragraph{LPP modes:} The LPP modes with $A_u$ and $B_u$ symmetry follow from Eqs.~(\ref{eq:bLPP}) and~(\ref{eq:acLPP}), respectively. The general solutions of these equations provide 5 LPP modes with $A_u$, and 10 LPP modes with $B_u$ symmetry. We found an isotropic plasma frequency parameter of $\omega_{\stext{p},x}=\omega_{\stext{p},y}=\omega_{\stext{p},z}=\omega_{\stext{p}}=1094.(8)$ cm$^{-1}$ sufficient to match all spectra $\varepsilon_{xx}$, $\varepsilon_{yy}$, and $\varepsilon_{zz}$ (Tab.~\ref{tab:fccepsDCinf}). This value is used to derive the LPP modes for our samples. We further assume that all samples investigated here share the same set of free charge carriers. This assumption is reasonable since both specimens were cut from the same bulk crystal. However, small gradients in Sn dopant volume density may exist throughout the bulk crystal due to the directional growth method and diffusion gradients near the solution solid interface~\cite{AidaEDFFGa2O3,ShimamuraAPPAGa2O3}. The resulting LPP mode frequencies are then summarized in Tab.~\ref{tab:LPPAuBuGSE}. 

\paragraph{LPP mode dispersion:} The LPP mode coupling for $A_u$ symmetry is trivial and equivalent to any other semiconductor material whose unit eigen displacement vectors are all parallel and/or orthogonal. Coupling for modes with $B_u$ symmetry is not trivial. Eq.~(\ref{eq:acLPP}) describes the LO plasmon coupling, and predicts the LPP mode frequencies  within a given sample as a function of the free charge carrier properties. For $\beta$-Ga$_2$O$_3$, the effective mass parameter anisotropy may need to be considered. Presently, available information suggest that the effective mass is nearly isotropic (see below). We therefore select to render the effects of free charge carriers by using an isotropic plasma frequency contribution, $\omega_{\stext{p}}$. We plot the resulting LPP modes with $B_u$ symmetry in Fig.~\ref{fig:BuLPP} as a function of $\omega_{\stext{p}}$. We also plot their unit eigen displacement vectors obtained from Eq.~(\ref{eq:acLPPalpha}) in Fig.~\ref{fig:BuLPPvec}.

A mode branch like behavior with phonon like and plasma like branches similar to orthogonal eigenvector lattice materials can be seen. For $\omega_p\rightarrow 0$, the upper LPP branches emerge from  LO mode frequencies, and the lowest 2 branches behave like uncoupled plasma modes. For $\omega_p\rightarrow\infty$, the 2 upper LPP branches behave like uncoupled plasma modes, and the lower branches behave like TO modes. Each LPP mode merges with one TO mode except for 2 high frequency plasma like branches. The unit eigen displacement vectors of the 2 plasma like modes approach the $x$ and $y$ directions for large plasma frequencies, and indicate a quasi orthorhombic free charge carrier response towards visible light optical frequencies. For intermediate $\omega_p$, the LPP coupling causes branch crossing with TO modes, which do not occur in orthogonal eigenvector lattice materials. The horizontal lines in Fig.~\ref{fig:BuLPP} indicate the $B_u$ symmetry TO modes.

\begin{table*}[!t]
\caption{LPP frequency parameters for $A_u$ and $B_u$ modes obtained from best match model analysis of $\varepsilon_{xx}$, $\varepsilon_{yy}$, $\varepsilon_{zz}$, and $\varepsilon_{xy}$. Also given are the eigenvector polarization angles $\alpha_{LPP}$ relative to $\mathbf{c}$. The last digit which is determined within the 90\% confidence interval is indicated with brackets.}
\begin{center}
\begin{tabular}{lcccccccccc}
    \noalign{\bigskip} \hline \hline\\
    Parameter &k=1&2&3&4&5&6&7&8&9&10\\
    \noalign{\smallskip} \hline
    $\omega_{\stext{LPP},k}$ [cm$^{-1}$] ($B_u$)& 967.(1)&872.(9) &730.(7)&638.(2)&458.(1)&357.(9)&331.(8)&277.(4)&226.(4)&188.(8)\\
    $\alpha_{\stext{LPP},k}$ [deg] ($B_u$)&179.(3) & 91.(1)&35.(4) &76.(8) & 160.(8)&36.(7)&91.(7) &67.(4) &100.(8) & 6.(7)\\
		$\omega_{\stext{LPP},k}$ [cm$^{-1}$] ($A_u$)&934.(8)&671.(9)&504.(7)&296.(8)&154.(6)& -& -& -& - &-  \\
		\hline \hline 
\end{tabular} \label{tab:LPPAuBuGSE}
\end{center}
\end{table*}

\paragraph{Free charge carrier properties:}

\begin{table}[!t]
\caption{Best match model parameters for free charge carrier contributions, static and high frequency dielectric constants. From our analysis we also obtain $\varepsilon_{\stext{DC},xy}=-0.0(5)$ and $\varepsilon_{\infty,xy}=-0.00(0)$ consistent with the generalized LST relation in Eq.~(\ref{eq:LSTxy}). Values reported from our analysis for $\varepsilon_{\stext{DC},zz}$ and $\varepsilon_{\infty,zz}$ are consistent with the traditional LST relation in Eq.~(\ref{eq:LST}) with TO and LO modes given in Tab.~\ref{tab:TOAuBuGSE}.}
\begin{center}
\begin{tabular}{lcccc}
    \noalign{\bigskip} \hline \hline
    & & $\varepsilon_{xx}$ ($\mathbf{c}$) & $\varepsilon_{yy}$ ($\mathbf{\hat{a}}$)& $\varepsilon_{zz}$($\mathbf{b}$)\\
    \noalign{\smallskip} \hline
    $\gamma_{\stext{p},(j)}$ [cm$^{-1}$]& this work&37(0).&46(0).&42(4).\\
   	$\mu_{(j)}$ [cm$^{2}$/(Vs)]&this work &8(3).&4(3).&8(8).\\
		$\varepsilon_{\infty,(j)}$& this work&3.8(9)&2.9(0)&3.8(7)\\
		$\varepsilon_{\stext{DC},(j)}$& this work&11.5(1)&11.8(9)&11.1(5)\\ \hline 
    $\varepsilon_{\infty,(j)}$& Theory Ref.~\cite{LiuAPL2007Ga2O3phononcalc} & 3.85&3.81&4.08 \\
		$\varepsilon_{\stext{DC},(j)}$& Theory Ref.~\cite{LiuAPL2007Ga2O3phononcalc} & 13.89 &10.84 & 11.49 \\
		$\varepsilon_{\infty,(j)}$& Theory Ref.~\cite{HePRB2006Ga2O3calc}&2.86&2.78& 2.84\\
		$\varepsilon_{\infty}$& Exp Ref.~\cite{RebienAPL2002Ga2O3SEfilms}& &3.57$^{a}$&  \\
		$\varepsilon_{\infty}$& Exp Ref.~\cite{PasslackJAP1995Ga2O3film}& &3.53$^{a}$&  \\
		$\varepsilon_{\stext{DC}}$& Exp Ref.~\cite{PasslackAPL1994Ga2O3films}& &9.9-10.2$^{a}$& \\
		$\varepsilon_{\stext{DC}}$& Exp Ref.~\cite{HoeneisenSSE1971Ga2O3DC}& &10.2$^{a}$& \\
		$\varepsilon_{\infty}$& Exp Ref.~\cite{SchmitzJAP1998Ga2O3films}& &3.6$^{a}$&  \\
		$\varepsilon_{\stext{DC}}$& Exp Ref.~\cite{SchmitzJAP1998Ga2O3films}& &9.57$^{a}$& \\
					\hline \hline 
& \multicolumn{4}{c}{$^a$Isotropic average from films.}\\ 
\end{tabular}
\end{center}\label{tab:fccepsDCinf}
\end{table}

Tab.~\ref{tab:fccepsDCinf} summarizes the Drude model parameters obtained from $\varepsilon_{xx}$, $\varepsilon_{yy}$, and $\varepsilon_{zz}$. For $\varepsilon_{xy}$ no significant Drude contribution was detected. In order to derive the free charge carrier density and mobility parameters from the plasma frequency and broadening parameters one needs the effective mass parameters. Unless magnetic fields are exploited and the optical Hall effect can be measured~\cite{SchubertJOSAA20_2003,HofmannAPL88_2006,HofmannAPL90_2007,HofmannJEM_2008,SchocheAPL98_2011,KuehnePRL111_2013,SchoecheAPL103_2013,KuehneRSI2014} long wavelength ellipsometry requires these parameters from auxiliary investigations. 

Experimental data on the electron effective mass in $\beta$-Ga$_2$O$_3$ is not exhaustive. Early estimates suggested 0.55~$m_e$~\cite{UeadaAPL1997Ga2O3}. A recent calculation predicts the effective electron mass at the $\Gamma$-point of the Brillouin zone almost isotropic with values between 0.27~$m_e$ and 0.28~$m_e$, depending on direction~\cite{Peelaerspssb2015Ga2O3meff}. These values agree with experimental measurements from angular resolved photoemission ARPES on $\mathbf{b^{\star}}$$\mathbf{c^{\star}}$-cleavage plane of (100) $\beta$-Ga$_2$O$_3$ (0.28~$m_e$, Ref.~\cite{MohamedAPL2010Ga2O3HRARPES,JanowitzNJP2011}). Earlier calculations using various approaches obtained 0.28~$m_e$~\cite{VarleyAPL2010Ga2O3DFT,HwangAPL2014Ga2O3FET}, 0.34~$m_e$~\cite{HePRB2006Ga2O3calc}, and 0.390~$m_e$~\cite{JuJMCA2014Ga2O3theory}. Calculations that did not use a hybrid functional approaches lead to smaller values of (0.23~$\dots$ 0.24)~$m_e$ in the local density approximation~\cite{YamaguchiSSC2004FPGa2O3} and (0.12~$\dots$0.13)~$m_e$ in the generalized gradient approximation~\cite{HeAPL2006Ga2O3theory}. He~\textit{et al.} reported slightly anisotropic electron effective mass values with $m_{\mathbf{a^{\star}}}$=0.123~$m_e$, $m_{\mathbf{c^{\star}}}$=0.124~$m_e$, and $m_{\mathbf{b^{\star}}}$=0.130~$m_e$, along axes $\mathbf{a^{\star}}$, $\mathbf{c^{\star}}$, and $\mathbf{b^{\star}}$, respectively, with ratios $m_{\mathbf{a^{\star}}}$/$m_{\mathbf{c^{\star}}}$=0.99 and $m_{\mathbf{b^{\star}}}$/$m_{\mathbf{c^{\star}}}$=1.05~\cite{HeAPL2006Ga2O3theory}. Yamaguchi also reported values with small anisotropy $m_{xx}$=0.2315~$m_e$ $m_{yy}$=0.2418$m_e$ $m_{zz}$=~$0.2270m_e$ using first principles full potential linearized augmented plane wave method~\cite{YamaguchiSSC2004FPGa2O3}. For analysis of the FIR-GSE and IR-GSE data we assume an isotropic effective electron mass value of 0.28~$m_e$, which appears to be a good compromise of the experimental and theoretical data. We then obtain $N=3.7(1)\times 10^{18}$ cm$^{-3}$, and anisotropic mobility parameters given in Tab.~\ref{tab:fccepsDCinf}. We observe similar mobility values along directions $x$ ($\mathbf{c}$) and $z$ ($\mathbf{b}$) and about 2 times smaller mobility along $\mathbf{y}$ ($\perp \mathbf{c},\mathbf{b}$).

\paragraph{Static and high frequency dielectric constant:} Tab.~\ref{tab:fccepsDCinf} also summarizes static and high frequency dielectric constants obtained in this work. We observe no significant contributions to $\varepsilon_{xy}$, both at $\omega \rightarrow 0$ and $\omega \rightarrow \infty$. At DC frequencies, $\beta$-Ga$_2$O$_3$ behaves quasi orthorhombic. We find that $\varepsilon_{\stext{DC},yy} (11.89) > \varepsilon_{\stext{DC},xx} (11.51) > \varepsilon_{\stext{DC},yy}$ (11.15), with very small anisotropy. In the high frequency limit, which is merely above the reststrahlen range for this work, $\beta$-Ga$_2$O$_3$ behaves nearly as an optically uniaxial crystal, with $\varepsilon_{\stext{DC},xx} (3.89) \approx \varepsilon_{\stext{DC},yy} (3.87) > \varepsilon_{\stext{DC},yy}$ (2.90). Data for the $x$-$y$ plane are consistent with the generalized LST relation in Eq.~(\ref{eq:LSTxy}), and for direction $z$ with Eq.~(\ref{eq:LST}). An isotropic average between all values obtained here is $\varepsilon_{\stext{DC}}$ = 11.52 and $\varepsilon_{\infty}$ = 3.53. A static dielectric constant between 9.9 and 10.2 was measured on films deposited by electron beam evaporation and annealing onto silicon and GaAs~\cite{PasslackJAP1995Ga2O3film}, and 10.2 was measured for single crystal $\beta$-Ga$_2$O$_3$ platelets in the direction perpendicular to the (100) plane at radio frequencies (5 kHz to 500 kHz)~\cite{HoeneisenSSE1971Ga2O3DC}. Schmitz, Gassmann, and Franchy report static and high frequency values from lineshape analysis of electron energy loss spectroscopy data from $\beta$-Ga$_2$O$_3$ films on metal substrates~\cite{SchmitzJAP1998Ga2O3films}. Values obtained previously for films agree well with our isotropic average~\cite{RebienAPL2002Ga2O3SEfilms,PasslackJAP1995Ga2O3film,SchmitzJAP1998Ga2O3films}, while previously reported isotropic DC values are slightly smaller~\cite{PasslackAPL1994Ga2O3films,HoeneisenSSE1971Ga2O3DC,SchmitzJAP1998Ga2O3films}. Data from recent band structure calculations are included in Tab.~\ref{tab:fccepsDCinf} and show some agreement with our results~\cite{LiuAPL2007Ga2O3phononcalc,HePRB2006Ga2O3calc}. Because it appears that our present work is the first comprehensive analysis of the long wavelength dielectric function tensor of single crystal $\beta$-Ga$_2$O$_3$ we believe that our data likely represent the most accurate values for this monoclinic semiconductor.

Finally, the effective monoclinic angles near DC and high frequencies (above the restrahlen range), according to Eqs.~(\ref{eq:monoclinicepsinfangle}) and~(\ref{eq:monoclinicepsxxDCangle}), respectively, approach 90$^{\circ}$ because $\varepsilon_{xy} \rightarrow 0$, both at $\omega \rightarrow 0$ and $\omega \rightarrow \infty$.

\section{Conclusions}

A dielectric function tensor model approach suitable for calculating the optical response of monoclinic and triclinic symmetry materials with multiple uncoupled long wavelength active modes was presented. The approach was applied to monoclinic $\beta$-Ga$_2$O$_3$ single crystal samples. Surfaces cut under different angles from a bulk crystal, (010) and ($\bar{2}$01), are investigated by generalized spectroscopic ellipsometry within infared and farinfrared spectral regions. We determined the frequency dependence of 4 independent $\beta$-Ga$_2$O$_3$ Cartesian dielectric function tensor elements by matching large sets of experimental data using a polyfit, wavelength by wavelength data inversion approach. From matching our monoclinic model to the obtained 4 dielectric function tensor components, we determined 4 pairs of transverse and longitudinal optic phonon modes with $A_u$ symmetry, and 8 pairs with $B_u$ symmetry, and their eigenvectors within the monoclinic lattice. We further report on density functional theory calculations on the infrared and farinfrared optical phonon modes, which  are in excellent agreement with our experimental findings. We derived and reported density and anisotropic mobility parameters of the free charge carriers within the tin doped crystals. We observed  5 longitudinal phonon plasmon coupled modes in $\beta$-Ga$_2$O$_3$ with $A_u$ symmetry and 10 modes with $B_u$ symmetry. We discussed and presented their dependence on an isotropic free charge carrier plasma. We also discussed and presented monoclinic dielectric constants for static electric fields and frequencies above the reststrahlen range, and we provided a generalization of the Lyddane-Sachs-Teller relation for monoclinic lattices with infrared and farinfrared active modes. We observed that the generalized Lyddane-Sachs-Teller relation is fulfilled excellently for $\beta$-Ga$_2$O$_3$. The model provided in this work will establish a useful base for infrared and farinfrared ellipsometry analysis of homo- and heteroepitaxial layeres grown on arbitrary faces of $\beta$-Ga$_2$O$_3$ substrates.

\section{Acknowledgments} This work was supported in part by the National Science Foundation (NSF) through the Center for Nanohybrid Functional Materials (EPS-1004094), the Nebraska Materials Research Science and Engineering Center (DMR-1420645) and awards CMMI 1337856 and EAR 1521428. We acknowledge further support from the Swedish Research Council (VR) under grant No.~2013-5580 and No.~2010-3848, the Swedish Governmental Agency for Innovation Systems (VINNOVA) under the VINNMER international qualification program, grant No. 2011-03486 and No. 2014-04712, and the Swedish Foundation for Strategic Research (SSF) under grant No.~FFL12-0181 and No.~RIF14-055. The financial support by the Link\"{o}ping Linnaeus Initiative on Nanoscale Functional Materials (LiLiNFM) supported by VR is gratefully acknowledged.  The authors further acknowledge grant support by the University of Nebraska-Lincoln, the J.~A.~Woollam Co., Inc., and the J.~A.~Woollam Foundation.

\bibliography{CompleteLibrary}

\end{document}